\documentclass[11pt,a4paper]{article}
\usepackage{amssymb}
\usepackage{epsfig}
\usepackage{cite}
\usepackage{array}
\usepackage[isu]{caption}
\usepackage{pennames}
\usepackage{textcomp}
\usepackage{rotating}
\usepackage{flafter}
\def\up{\rule[0mm]{0mm}{3ex}}
\def\down{\rule[-1.5ex]{0mm}{3ex}}
\def\both{\rule[-1.5ex]{0mm}{4.5ex}}
\newcommand{\Pgtm}      {\ensuremath{\mathrm{\tau^-}}}

\newcommand{\KOspp}   {\PKzS\rightarrow\Pgpp\Pgpm}
\newcommand{\KOsppn}  {\PKzS\rightarrow\Pgpz\Pgpz}

\newcommand{\Km}        {(\PK)^-}
\newcommand{\Kpm}       {(\PK \Pgp)^-}
\newcommand{\KpKetam}   {(\PK \Pgp+\PK \Pgh)^-}
\newcommand{\Ketam}     {(\PK \Pgh)^-}
\newcommand{\Kppm}      {(\PK \Pgp \Pgp)^-}
\newcommand{\KppKetapm} {(\PK \Pgp \Pgp + \PK \Pgh \Pgp)^-}
\newcommand{\Ketapm}    {(\PK \Pgh \Pgp)^-}
\newcommand{\Kpppm}     {(\PK \Pgp \Pgp \Pgp)^-}

\newcommand{\tauK}       {\Pgtm \rightarrow \PKm \Pgngt}
\newcommand{\tauppn}     {\Pgtm \rightarrow \Pgpm \Pgpz \Pgngt}
\newcommand{\tauKpn}     {\Pgtm \rightarrow \PKm \Pgpz \Pgngt}
\newcommand{\tauKOp}     {\Pgtm \rightarrow \PKz \Pgpm \Pgngt}
\newcommand{\tauKSp}     {\Pgtm \rightarrow \PKzS \Pgpm \Pgngt}
\newcommand{\tauKLp}     {\Pgtm \rightarrow \PKzL \Pgpm \Pgngt}
\newcommand{\tauKpnpn}   {\Pgtm \rightarrow \PKm \Pgpz \Pgpz \Pgngt}
\newcommand{\tauKpppn}   {\Pgtm \rightarrow \PKm \Pgpp \Pgpm \Pgpz \Pgngt}
\newcommand{\tauKppNpn}  {\Pgtm \rightarrow \PKm \Pgpp \Pgpm (\mbox{n}\Pgpz) \Pgngt}

\newcommand{\tauKppOpnex}{\Pgtm \rightarrow \PKm \Pgpp \Pgpm (0\Pgpz, \mathrm{ex. \PKz})\Pgngt}
\newcommand{\tauKeta}    {\Pgtm \rightarrow \PKm \Pgh \Pgngt}

\newcommand{\tauKSetapn} {\Pgtm \rightarrow \PKst \Pgh \Pgpz \Pgngt}

\newcommand{\tauppp}     {\Pgtm \rightarrow \Pgpm \Pgpm \Pgpp \Pgngt}

\newcommand{\tauKpnpnpn} {\Pgtm \rightarrow \PKm \Pgpz \Pgpz \Pgpz \Pgngt}
\newcommand{\tauKpp}     {\Pgtm \rightarrow \PKm \Pgpp \Pgpm \Pgngt}
\newcommand{\tauKKp}     {\Pgtm \rightarrow \Pgpm \PKm \PKp \Pgngt}

\newcommand{\tauKKO}     {\Pgtm \rightarrow \PKm \PKz \Pgngt}
\newcommand{\tauKOKOp}   {\Pgtm \rightarrow \PKz \PaKz \Pgpm \Pgngt}
\newcommand{\tauKOppn}   {\Pgtm \rightarrow \PKz \Pgpm \Pgpz \Pgngt}
\newcommand{\tauKOKpn}   {\Pgtm \rightarrow \PKz \PKm \Pgpz \Pgngt}
\newcommand{\tauKOppnpn} {\Pgtm \rightarrow \PKz \Pgpm \Pgpz \Pgpz \Pgngt}

\newcommand{\tauKOppp}   {\Pgtm \rightarrow \PKz \Pgpm \Pgpp \Pgpm \Pgngt}

\newcommand{\K}       {\PKm \Pgngt}
\newcommand{\Kpn}     {\PKm \Pgpz \Pgngt}

\newcommand{\Kpnpn}   {\PKm \Pgpz \Pgpz \Pgngt}
\newcommand{\Kpppn}   {\PKm \Pgpp \Pgpm \Pgpz \Pgngt}
\newcommand{\KppNpn}   {\PKm \Pgpp \Pgpm \mbox{n} \Pgpz \Pgngt}
\newcommand{\Keta}    {\PKm \Pgh \Pgngt}
\newcommand{\Ketap}   {\PKm \Pgh \Pgpz \Pgngt}

\newcommand{\ppn}     {\Pgpm \Pgpz \Pgngt}
\newcommand{\ppnpn}   {\Pgpm \Pgpz \Pgpz \Pgngt}
\newcommand{\ppp}     {\Pgpm \Pgpp \Pgpm \Pgngt}
\newcommand{\ppppn}   {\Pgpm \Pgpp \Pgpm \Pgpz \Pgngt}

\newcommand{\Kpnpnpn} {\PKm \Pgpz \Pgpz \Pgpz \Pgngt}
\newcommand{\Kpp}     {\PKm \Pgpp \Pgpm \Pgngt}
\newcommand{\KKp}     {\Pgpm \PKm \PKp \Pgngt}

\newcommand{\KOp}     {\PKz \Pgpm \Pgngt}
\newcommand{\KKO}     {\PKm \PKz \Pgngt}
\newcommand{\KKOpn}   {\PKm \PKz \Pgpz \Pgngt}

\newcommand{\KOppn}   {\PKz \Pgpm \Pgpz \Pgngt}
\newcommand{\KOppnpn} {\PKz \Pgpm \Pgpz \Pgpz \Pgngt}
\newcommand{\pKOpp}   {\PKz \Pgpm  \Pgpz \Pgpz \Pgngt}
\newcommand{\KOppp}   {\PKz \Pgpm \Pgpp \Pgpm \Pgngt}

\newcommand{\as}      {\ensuremath{\alpha_\mathrm{s}}}

\newcommand{\dedx}   {\ensuremath{\mathrm{d}E/\mathrm{d}x}}

{\catcode`\.=\active\gdef.{\egroup\setbox2=\hbox\bgroup}}
\def\centerdots{\catcode`\.=\active\setbox0=\hbox\bgroup}
\def\endcenterdots{\egroup \ifvoid2 \setbox2=\hbox{~}\fi
   \ifdim \wd0>\wd2 \setbox2=\hbox to\wd0{\unhbox2\hfill}\else
     \setbox0=\hbox to\wd2{\hfill\unhbox0}\fi
   \catcode`\.=12 \box0.\box2}
\newcolumntype{C}{>{\centerdots}c<{\endcenterdots}}
\begin{document}
\begin{titlepage}
\begin{center}{\large    EUROPEAN ORGANIZATION FOR NUCLEAR RESEARCH
}\end{center}\bigskip
\begin{flushright}
CERN-PH-EP/2004-010 \\ 18. February 2004
\end{flushright}
\bigskip\bigskip\bigskip\bigskip\bigskip
\begin{center}
{\huge\bf\boldmath Measurement of the Strange Spectral Function in
  Hadronic \Pgt\ Decays} \\ 
\end{center}\bigskip\bigskip
\begin{center}{\LARGE The OPAL Collaboration}
\end{center}\bigskip\bigskip
\bigskip\begin{center}{\large Abstract}\end{center}
\noindent
Tau lepton decays with open strangeness in the final state are
measured with the OPAL detector at LEP to determine the strange
hadronic spectral function of the \Pgt\ lepton. The decays \Pgtm
$\rightarrow$ $\Kpm \Pgngt$, $\Kppm  \Pgngt$ and $\Kpppm  \Pgngt$ with
final states consisting of neutral and charged kaons and pions have
been studied. The invariant mass distributions of
$93.4\%$  of these final states have been experimentally
determined. Monte Carlo simulations have been used for the remaining
$6.6\%$ and for the strange final states including \Pgh\ mesons. 
The reconstructed strange final states, corrected for resolution
effects and detection efficiencies, yield the strange spectral
function of the \Pgt\ lepton. The moments of the spectral function and the
ratio of strange to non-strange moments, which are important
input parameters for theoretical analyses, are determined. 
Furthermore, the branching fractions 
$B(\tauKpn)=(0.471\pm0.064_\mathrm{stat}\pm0.022_\mathrm{sys})\%$
and 
$B(\tauKpp)=(0.415\pm0.059_\mathrm{stat}\pm0.031_\mathrm{sys})\,\%$
have been measured.
\bigskip
\begin{center}{(\em\large To be submitted to Euro.Phys.J. C)}\end{center}
\end{titlepage}

\pagestyle{empty}
\vspace*{-1cm}
\begin{center}{\Large        The OPAL Collaboration
}\end{center}\bigskip
\begin{center}{
G.\thinspace Abbiendi$^{  2}$,
C.\thinspace Ainsley$^{  5}$,
P.F.\thinspace {\AA}kesson$^{  3,  y}$,
G.\thinspace Alexander$^{ 22}$,
J.\thinspace Allison$^{ 16}$,
P.\thinspace Amaral$^{  9}$, 
G.\thinspace Anagnostou$^{  1}$,
K.J.\thinspace Anderson$^{  9}$,
S.\thinspace Arcelli$^{  2}$,
S.\thinspace Asai$^{ 23}$
% ...
}\end{center}\bigskip
\bigskip
$^{  1}$School of Physics and Astronomy, University of Birmingham,
Birmingham B15 2TT, UK
\newline
$^{  2}$Dipartimento di Fisica dell' Universit\`a di Bologna and INFN,
I-40126 Bologna, Italy
\newline
$^{  3}$Physikalisches Institut, Universit\"at Bonn,
D-53115 Bonn, Germany
\newline
% ...
$^{  5}$Cavendish Laboratory, Cambridge CB3 0HE, UK
\newline
% ...
$^{  9}$Enrico Fermi Institute and Department of Physics,
University of Chicago, Chicago IL 60637, USA
\newline
% ...
$^{ 16}$Department of Physics, Schuster Laboratory, The University,
Manchester M13 9PL, UK
\newline
% ...
$^{ 22}$Department of Physics and Astronomy, Tel Aviv University,
Tel Aviv 69978, Israel
\newline
$^{ 23}$International Centre for Elementary Particle Physics and
Department of Physics, University of Tokyo, Tokyo 113-0033, and
Kobe University, Kobe 657-8501, Japan
\newline
% ...
$^{  y}$ now at CERN
\newline
% ...
\clearpage
\pagestyle{plain}
\section{Introduction}\label{sec:Intr}

The \Pgt\ lepton is the only lepton heavy enough to decay into hadrons. A
comparison of the inclusive hadronic decay rate of the \Pgt\ lepton
with QCD predictions allows the measurement of some fundamental
parameters of the theory. The inputs to these studies are the spectral
functions that measure the transition probability to create hadronic
systems of invariant mass squared $s=m^2$.

The energy regime accessible in \Pgt\ lepton decays can be divided
into two different regions: the low energy regime which has a rich
resonance structure  where non-perturbative QCD dominates; and the
high energy regime near the kinematic limit,
$s=m_\Pgt^2=(1.777\,\mathrm{GeV})^2$, where perturbative QCD
dominates. The high energy regime in \Pgt\ lepton decays provides an
environment where the strong coupling constant $\alpha_s$ can be measured
\cite{bib:Braaten88,bib:Braaten89,bib:Narison88,bib:Braaten,bib:Pivovarov1991rh,bib:Pich92}
because the perturbative expansion converges well and non-perturbative
effects are small. 
The measurement of the non-strange spectral function of hadronic \Pgt\
lepton decays \cite{bib:OPALSven,bib:ALEPHQCD,bib:CLEOQCD}
has provided one of the most accurate measurements of \as, and
some very stringent tests of perturbative QCD at  
relatively low mass scales \cite{bib:TheoAlphas}.
The spectral function of strange decays allows additional and
independent tests of QCD and a measurement of the mass of the strange
quark
\cite{bib:Chetyrkin,bib:Pich1999hc,bib:Maltman2002wb,bib:Braaten}. 

\begin{table}[b!]
  \begin{center}
    \small{
    \begin{tabular}{r|c||l|c|l|c}\hline
      \multicolumn{2}{c||}{\both ~} & \multicolumn{2}{|c|}{Measured in
      this Paper} & \multicolumn{2}{|c}{Not Measured} \\ 
      \multicolumn{1}{r|}{\both ~} & $B_\mathrm{total}/\%$ & \Pgt\ decay & $B_{\mathrm{PDG}}/\%$ & \Pgt\ decay & \multicolumn{1}{c}{$B_{\mathrm{PDG}}/\%$} \\\hline\hline
      \both
      $\Km$         & $0.686\pm 0.023$   & ~                   & ~                 & $\tauK$            &     $0.686\pm 0.023$   \\\cline{3-6}
      \up
      $\Ketam$      & $0.027\pm 0.006$   & ~                   & ~                 & $\tauKeta$            &     $0.027\pm 0.006$   \\
      $\Kpm$        & $1.340\pm0.050$   & $\tauKpn$           & $0.450\pm0.030$  & ~                  &     ~   \\
      \down ~                & ~         & $\tauKOp$           & $0.890\pm0.040$  & ~                  &     ~   \\\cline{3-6}
      \up
      $\Kppm$      & $0.708\pm0.068$ & $\tauKOppn$      & $0.370\pm0.040$  & ~                                    &     ~   \\
      ~                & ~              & $\tauKpp$          & $0.280\pm0.050$  & ~                                    &     ~   \\
      ~          & ~              & ~                                    & ~                 & $\tauKpnpn$      & $0.058\pm 0.023  $ \\
      \down $\Ketapm$    & $0.029\pm 0.009$   & ~                   & ~                 & $\tauKSetapn$            &     $0.029\pm 0.009$   \\\cline{3-6}
      \cline{3-6}
      \up
      $\Kpppm$                   & $0.150\pm0.045$  & $\tauKpppn$   & $0.064\pm0.024$ & ~& ~ \\
            ~                    & ~                 & ~           &       ~              & $\tauKOppnpn$ & $0.026\pm0.024$ \\
            ~                    & ~                 & ~           &       ~              & $\tauKpnpnpn$ & $0.037\pm0.021$ \\
       \down~                    & ~                 & ~           &       ~              & $\tauKOppp$   & $0.023\pm0.020$ \\
      \hline\hline
      \both\mbox{Sum}  & $2.940\pm 0.099$ & ~                                    & $2.054\pm 0.085$  & ~                  & $0.886\pm0.051$

    \end{tabular}}
    \caption[Measured strange channels]
    {\em Overview of all channels with net strangeness and their
    branching fractions \cite{bib:PDG}.}
    \label{tab:channels}
  \end{center}
\end{table}

The strange spectral function of the \Pgt\ lepton is obtained from
the invariant mass spectra of hadronic \Pgt\ decay modes with net
strangeness. The relevant decay channels are listed in Table
\ref{tab:channels}.   
The only contribution to the one meson final state\footnote{In order
  to simplify the text we refer only to the decays of the negatively
  charged \Pgt\ lepton. Simultaneous treatment of the charge conjugate decay
  is always implied.} $\tauK$ is not reconstructed in this
analysis. $\Kpm$ and $\Ketam$ contribute to the two-meson final
states. They have a total branching fraction of $1.367\%$. The decay
{$\tauKpn$} with a branching fraction of $0.450\%$ can be fully
reconstructed. The {$\tauKOp$} channel ($B=0.890\%$) consists of
two decay chains: $\tauKLp$ ($50\%$) and $\tauKSp$ ($50\%$) where
the \PKzS\ decays to two charged (neutral) pions in roughly $68\%$
($32\%$) of all cases. Only those final states
where the \PKzS\ decays to two charged pions are considered. The
same applies to {$\tauKOppn$} ($B=0.370\%$). This and
{$\tauKpp$} $(B=0.280\%)$ together form the two most important
contributions to the three-meson final states. Finally, the
four-meson final state $\Kpppm$ is detected via the decay
$\tauKpppn$. In addition, the decay \Ketap\ contributes with
$0.029\%$. Hence, $93.4\%$ of all decay channels of the
multi-meson final states with open strangeness, $\Kpm$, $\Kppm$ and 
$\Kpppm$, were measured. This paper describes the selection of these 
dominant channels and the measurement of their invariant mass
spectra using data collected with the OPAL detector during the LEP-I
period from 1991 to 1995. The remaining $6.6\%$ and the final states
including \Pgh\ mesons are taken from Monte Carlo simulation. The
spectral function is then determined from these spectra and the
spectral moments are calculated.

From an experimental point of view, one of the key issues of this
analysis is the separation of charged kaons and pions via the
measurement of energy loss in the OPAL jet chamber in the dense
environment of multiprong \Pgt\ lepton decays. Substantial
improvements have been achieved compared to previous publications
\cite{bib:smjt}. In particular, these improvements have made it
possible to obtain a reliable \dedx\ measurement in an environment where
three tracks are very close to each other.  
The reconstruction of neutral pions is based on the study of shower
profiles in the electromagnetic calorimeter. Furthermore the
identification and reconstruction of \Pgt\ lepton decays with
{$\KOspp$} has been achieved with high efficiency and good mass
resolution.   

The outline of this paper is as follows:
Section \ref{sec:OPDe} gives a short description of the OPAL detector
concentrating on those components which are important for this
analysis. In addition, the \Pgt\ lepton selection and the Monte Carlo
samples used are discussed. Section \ref{sec:ExAs} continues with a
discussion of the experimental aspects of this work. The selection of
the strange hadronic \Pgt\ lepton decays is described in Section
\ref{sec:SelEv}. In Section \ref{sec:Res}, the results for the
branching fractions,  the strange spectral function and the spectral
moments are presented together with a discussion of the systematic
uncertainties. The results are summarized in Section \ref{sec:Summary}.

\section{Detector and Data Samples}\label{sec:OPDe}
\subsection{The OPAL Detector}
A detailed description of the OPAL detector can be found elsewhere
\cite{bib:Detector}. A short overview is given here of those
components that are vital for this analysis. 
Charged particles are tracked in the central detector, which is
enclosed by a solenoidal magnet, providing an axial magnetic field of
$0.435\,\mathrm{T}$. A high-precision silicon micro-vertex detector
surrounds the beam pipe. It covers the angular region of $|
\cos\theta |\leq 0.8$ and provides tracking information in the
$r-\phi$ direction\footnote{In the OPAL
  coordinate system the $x$-axis points to the center of the LEP
  ring. The $z$-axis is in the $e^-$ beam direction. The angle
  $\theta$ is defined relative to the $z$-axis and $\phi$ is the
  azimuthal angle with respect to the $x$-axis.} (and $z$ from 1993)
\cite{bib:DetectorSI}.  The silicon detector is surrounded by   
three drift chambers: a high-resolution vertex detector, a
large-volume jet chamber and $z$-chambers.

The jet chamber measures the momentum and energy loss of charged
particles over $98\,\%$ of the solid angle. It is subdivided into 24
sectors in $r-\phi$, each containing a radial plane with 159 anode
sense wires parallel to the beam pipe. Cathode wire planes form the
boundaries between adjacent sectors. The 3D-coordinates of points
along the trajectory of a track are determined from the sense wire
position, the drift time ($r-\phi$) and a charge division measurement
($z$) on the sense wire. The combined momentum resolution of the OPAL
tracking system is $\sigma_p/p^2\approx 1.5\cdot
10^{-3}\,\mathrm{GeV}^{-1}$. From the total charge on each anode wire,
the energy loss \dedx\ is calculated and used for particle
identification. This measurement provides a separation between pions
and kaons of at least $2\sigma$ in the momentum range relevant for
this analysis ($3\,\mathrm{GeV}<p<35\,\mathrm{GeV}$).    

Outside the solenoid are scintillation counters which measure the
time-of-flight from the interaction region and aid in the rejection of
cosmic events.  Next is the electromagnetic calorimeter (ECAL), which,
in the barrel section, is composed of $9440$ lead-glass blocks,
approximately pointing to the interaction region, and covering the
range $|\cos\theta|<0.82$. Each block has a $(10 \times 10)
\,\mathrm{cm}^2$ profile with a depth of $24.6$ radiation lengths. The
resolution of the ECAL in the barrel region, including the effects of
the approximately one radiation length of material in front, is
$\sigma_E/E=\sqrt{(0.16)^2 \mathrm{GeV}/E + (0.015)^2}$.

The hadron calorimeter (HCAL) is beyond the electromagnetic
calorimeter and is instrumented with layers of limited streamer tubes
in the iron of the solenoid magnet return yoke.  
The outside of the hadron calorimeter is surrounded by the muon chamber
system, which is composed of four layers of drift chambers in the
barrel region. 

\subsection{Selection of \Pgt\ Lepton Candidates}
For the selection of \Pgt\ lepton candidates, the standard \Pgt\
selection procedure described in~\cite{bib:TPsel} is used. The decay
of the \PZz\ produces a pair of back-to-back highly relativistic \Pgt\
leptons. Their decay products are strongly collimated and well
contained within cones of half-angle $35^\circ$. Therefore, each \Pgt\
decay is treated separately. In order to have a precise and reliable
\dedx\ measurement and to avoid regions of non-uniform calorimeter
response, this analysis is restricted to the region
$|\cos\theta|<0.68$. To reject background from hadronic events,
exactly two cones are required and a maximum of six good\footnote{A
  good track has a minimum number of 20 hits in the jet chamber, a
  maximum $|d_0|$ of $2\,\mathrm{cm}$, a maximum $|z_0|$ of
  $75\,\mathrm{cm}$, at least $100\,\mathrm{MeV}$ transverse momentum
  and a maximum radius of the first measured point on the track of
  $75\,\mathrm{cm}$.} tracks in the event is  
allowed. This background is further reduced by requiring that the sum of the
charges of all tracks in each individual cone is $\pm 1$ and the net
charge of the whole event is zero.
A total of $162\,477$ \Pgt\ cone candidates survive these selection
criteria with an estimated non-\Pgt\ background fraction of $1.5\,\%$. 

\subsection{Simulation of Events}\label{sub:Smlt}
The \Pgt\ Monte Carlo samples used consist of $200\,000$ \Pgt\ pair events
generated at $\sqrt{s}=m_{\PZz}$ using KORALZ~4.02\cite{bib:KoralZ} and
a modified version of TAUOLA~2.4\cite{bib:Tauola}.
Modifications were necessary because all four- and five-meson final
states with kaons (signal as well as background) are missing in the
standard version. Since the resonance structure of these channels is
poorly known, only phase space distributions of these final states
were generated. In addition, the resonance structure of various final
states was modified to give a better description of the data
\cite{bib:Finkemeier1995sr,bib:Finkemeier1996hh}.  The branching
fractions of the decay channels with kaons are enhanced in this sample
so that it comprises roughly a factor of ten more \Pgt\ decays with kaons
than expected from data. The Monte Carlo events are then reweighted
to the latest branching fractions given in \cite{bib:PDG}, which are
used throughout the selection procedure. The Monte Carlo events were
processed through the GEANT 
OPAL detector simulation\cite{bib:GOPAL}.  

The non-\Pgt\ background was simulated using Monte Carlo samples that consist
of $4\,000\,000\,\mathrm{q}\bar\mathrm{q}$ events generated with
JETSET\cite{bib:Jetset}, $574\,000$ Bhabha events generated with BHWIDE
\cite{bib:Bhwide}, $792\,000$ $\mu$-pair events generated with
{KORALZ}\cite{bib:KoralZ} and $1\,755\,000$ two-photon events using
PHOJET\cite{bib:PHOJET}, F2GEN\cite{bib:TWOGEN}  and
VERMASEREN\cite{bib:Vermaseren,bib:Vermaseren2}.

\section{Identification of Hadrons in \Pgt\ Final States}\label{sec:ExAs}
\subsection{Energy Loss Measurement in \Pgt\  Decays with Three or More Tracks}
\label{sub:dedx}
A crucial part of this analysis is the identification of
charged kaons via energy loss measurement in the jet chamber. Since
this is the only means of distinguishing between charged pions and
kaons in OPAL, a very good understanding of the effects present in the
multi-track environment in \Pgt\ lepton decays is vital for any
analysis that requires particle identification.  

The high Lorentz boost $(\gamma\approx 25)$ of the \Pgt\ lepton results
in its decay products being contained in a narrow cone with a typical
opening angle of $5^\circ$. In those cases where the final state
consists of more than one track, the \dedx\ measurement is known to
be no longer reliable \cite{bib:smjt}. A systematic shift in the
\dedx\ distribution is observed which leads to a misidentification of
charged pions as kaons and thus to a reduced sensitivity in those
cases where particle identification is required. The reason is
explained in the following text.  

When a charged particle passes through the jet chamber, it produces an
ionization cloud. This cloud travels at a constant speed of
$v_D\approx 53\,$\textmu$\mathrm{ m/ns}$  in the homogeneous drift
field to the sense wires, where it produces a signal pulse. The pulse
is integrated over a time $\Delta t=200\,\mathrm{ns}$, which corresponds to a
drift distance of $\sim 1\,\mathrm{cm}$. The integral is proportional
to the energy loss of the particle in this particular drift volume. 
The hits used for the energy loss measurement are the same as used in the
track reconstruction, but they have to fulfill additional quality
criteria \cite{bib:OPALdEdx}. In the following, they are called \lq
\dedx\ hits\rq. 

\begin{figure}[!b]
  \begin{center}
    \includegraphics[width=.98\textwidth]{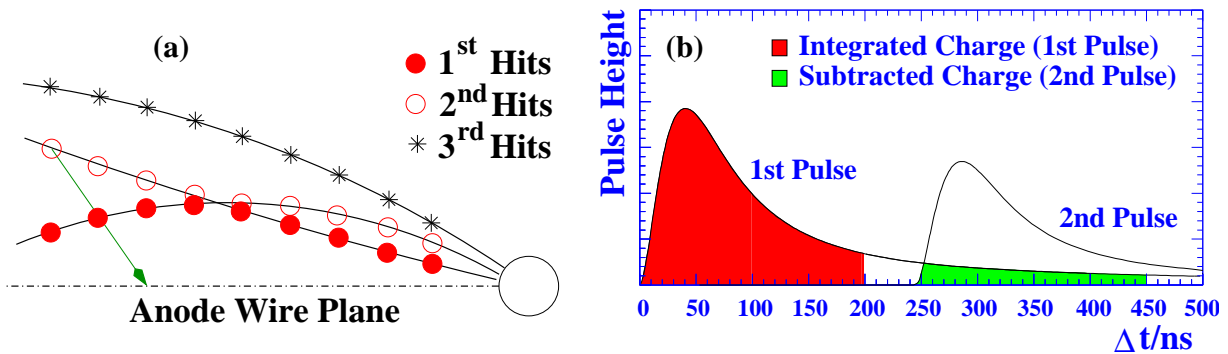}
    \caption{\em Illustration of the tail subtraction procedure. (a)
      shows one half-sector of the OPAL central drift
      chamber with three tracks. The arrow indicates the drift
      direction of the ionization cloud following a path given by a
      Lorentz angle of $20^\circ$; (b) illustrates the
      signal seen on the sense wire for two successive pulses.} 
    \label{fig:pulses}
  \end{center}
\end{figure}

\begin{figure}[!t]
  \begin{center}
    \includegraphics[width=.9\textwidth]{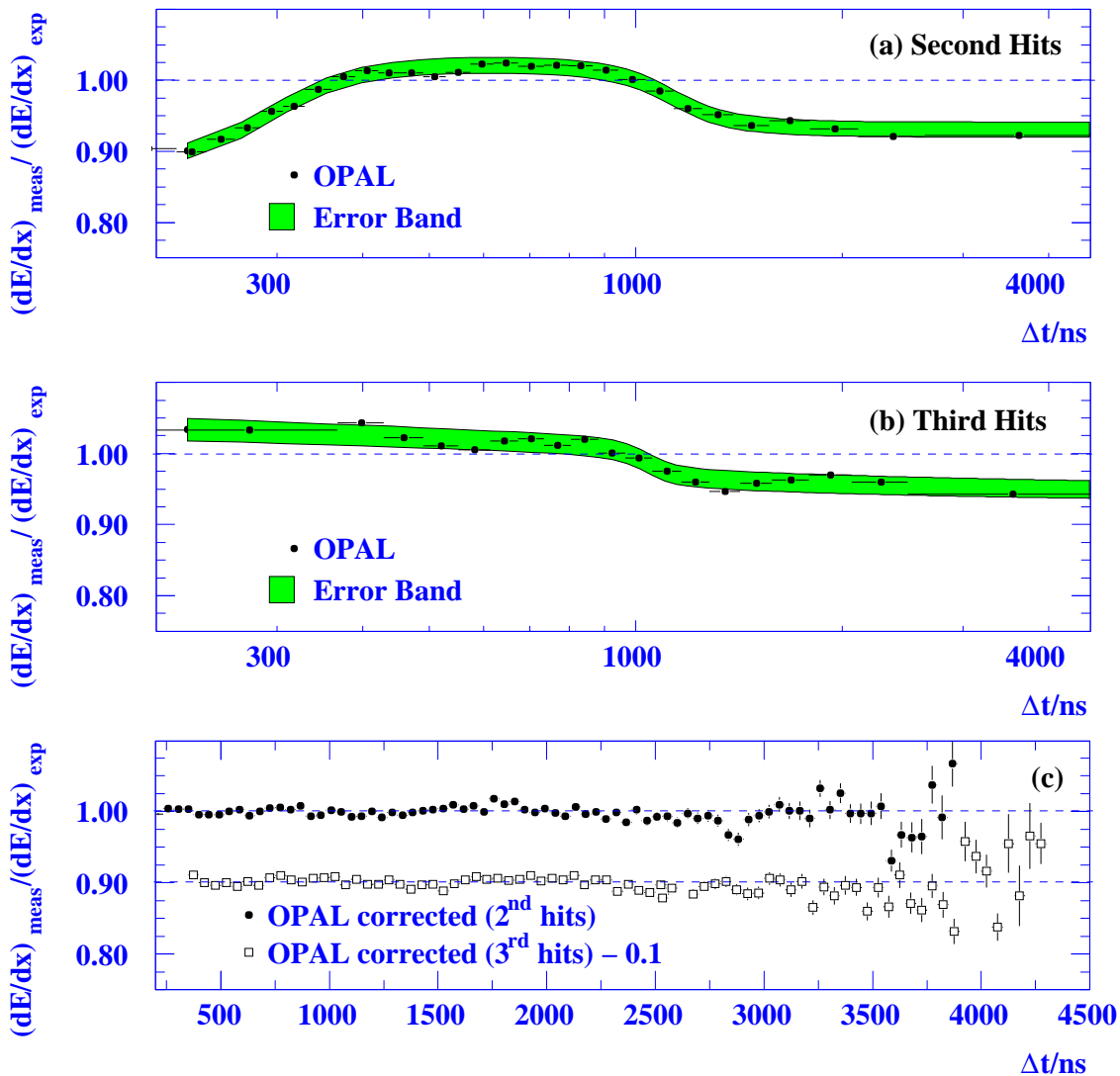}
    \caption{\em Measured energy loss normalized to the expectation as a
      function of time difference between two measured hits
      ($200\,\mathrm{ns} \mathrel{\widehat{=}}
      1.054\,\mathrm{cm}$). The observed deviations using the standard
      correction are shown for (a) \lq second hits\rq\ and (b) \lq
      third hits\rq. The error band reflects the $1\sigma$ error band
      of the parametrization. (c) shows the same distribution after
      all corrections have been applied. For better visibility, the
      distribution for \lq third hits\rq\ is shifted by $-0.1$ in the
      plot.}    
    \label{fig:correction_1}
  \end{center}
\end{figure}

If an additional track passes through the same sector, a second pulse will be
created which overlaps with the tail of the first one (Figure
\ref{fig:pulses}). The contribution of the tail is
determined by extrapolation of a reference pulse \cite{bib:OPALdEdx}
which is normalized to the integral over the first pulse.
The pulse used in all previous OPAL publications overestimates this
tail contribution. The observed deviation in the energy loss
measurement as a function of the distance between these two pulses
($\Delta t$) is shown in Figure \ref{fig:correction_1}. This plot is
obtained using tracks in \Pgt\ data that have \lq first hits\rq\ and \lq
second hits\rq, and which have a momentum greater than 
$3\,\mathrm{GeV}$. Such tracks arise when a second track traverses the
same sector of the drift chamber and the two tracks have at least one 
intersection point in $r-\phi$ within the sensitive volume. Thus, for
a given track, some of the hits do not need tail subtraction (\lq
first hits\rq) and some hits do need tail subtraction (\lq second
hits\rq). Using \lq first hits\rq\ only (given there are at least 20
of them) $(\dedx)_\mathrm{exp}$ is calculated. The \lq second hits\rq\
of the same track were then used to obtain $(\dedx)_\mathrm{meas}$
which is analyzed as a function of the drift time difference $\Delta
t$ between two successive hits. The left-right ambiguity, which in
principle exists if two tracks pass on either side of the anode plane,
is treated properly here, since the correction is applied based on
measured drift times and not on the spatial separation of the tracks.  

For two hits as close as $\Delta t=200\,\mathrm{ns}$, the
observed deviation is of the order of $10\%$ of the measured
\dedx. In the region between $400\,\mathrm{ns}<\Delta
t<900\,\mathrm{ns}$ the measured \dedx\ is slightly
overestimated. The standard correction is finally switched off for
pulses with a drift time difference of more than $1000\,\mathrm{ns}$
which produces the  structure shown in Figure
\ref{fig:correction_1}(a). The deviation as observed for \lq third
hits\rq\ can be seen in Figure \ref{fig:correction_1}(b). When weighted
with the $\Delta t$ distribution of all measured hits this gives on
average the correct \dedx\ for high multiplicity events. In
multiprong \Pgt\ lepton decays this $\Delta t$ distribution differs from the
average in the sense that the distribution peaks at smaller $\Delta
t$. Therefore, the energy loss measurement tends to be smaller than it
should be. To avoid this problem, previous analyses\cite{bib:smjt}
exploited the \dedx\ information only of those tracks closest to the
anode plane to classify the \Pgt\ decay mode. This however
significantly reduces the number of identified decays.    

For this analysis, a new reference pulse has been developed that
avoids the shortcomings of the standard one. In addition, a
parametrized pulse shape was used instead of a binned one to avoid
artifacts like the dip at $\Delta t\approx 500\,\mathrm{ns}$. The new
reference pulse is of the form  
\begin{eqnarray}
  P_\mathrm{norm} = &~& \left( p_1 \Delta t \exp(-\frac{\Delta t}{p_2}) 
   + p_3 (\Delta t)^2 \exp(-\frac{(\Delta t)^2}{p_4}) \right)\nonumber\\
   &+& \left( p_5 \Delta t \exp(-\frac{\Delta t}{p_6}) 
   + p_7 (\Delta t)^2 \exp(-\frac{(\Delta t)^2}{p_8})\right)\nonumber\\[1.5ex]
       &+& \tilde p_1 + \tilde p_2 \Delta t
\end{eqnarray}
with two terms to describe the short-range and the long-range part
respectively plus a linear contribution. The $p_i$ are parameters that
are optimized for the multi-track environment in \Pgt\ lepton
decays. The correction is applied to all hits. The effect of the
correction described above can be seen in Figure
\ref{fig:correction_1}(c) for \lq second hits\rq\ and \lq third
hits\rq. Apart from this normalization correction, a further bias
reduction is obtained by also correcting the shape of the reference
pulse depending on the charge deposited by the preceding pulse.  In
general, the new reference pulse shows a steeper rise at low $\Delta
t$ and a lower tail to avoid overestimation of the tail subtraction
for subsequent pulses as  explained above. As a result of this
procedure, a \dedx\ bias reduction to $\pm1\%$ in
$(\dedx)_\mathrm{meas}/(\dedx)_\mathrm{exp}$ has been achieved. As the
procedure is applied iteratively, i.e.~the first pulse is used to
correct the second, the first and corrected second pulse are used to
correct a possible third pulse and so on, the chosen reference pulse
is valid for any jet topology \cite{bib:OPALdEdx}.   

If a track is close to the anode or cathode plane
in the jet chamber, the drift field is no longer homogeneous and one observes a
deviation in the measured {\dedx} of the order of $3\%$. Corrections for this
effect are determined using $\PZz\to\Pgmm\Pgmp$ events. 

When a signal is measured at a sense wire, an induced signal at the
neighboring wires is also present. This effect depends on the momentum
(or curvature) of the track and on $\cos\theta$. Corrections have been
determined using $\mu$-pairs and $\tau\to\mu\bar\nu_\mu\nu_\tau$
decays. The effect is largest $(\sim 5\%)$ for $\cos\theta\sim 0$ and
large track momenta. 

Finally, hits are discarded from tracks where the corresponding hit from the
following or the preceding track is missing. In those cases, the measured
{\dedx} is overestimated since the charge is not correctly distributed among
the hits but is assigned to one hit only. By discarding this kind of
hits, $(3-5)\%$ of all {\dedx} hits are lost. 

\begin{figure}[!t]
  \begin{center}
    \includegraphics[width=.95\textwidth]{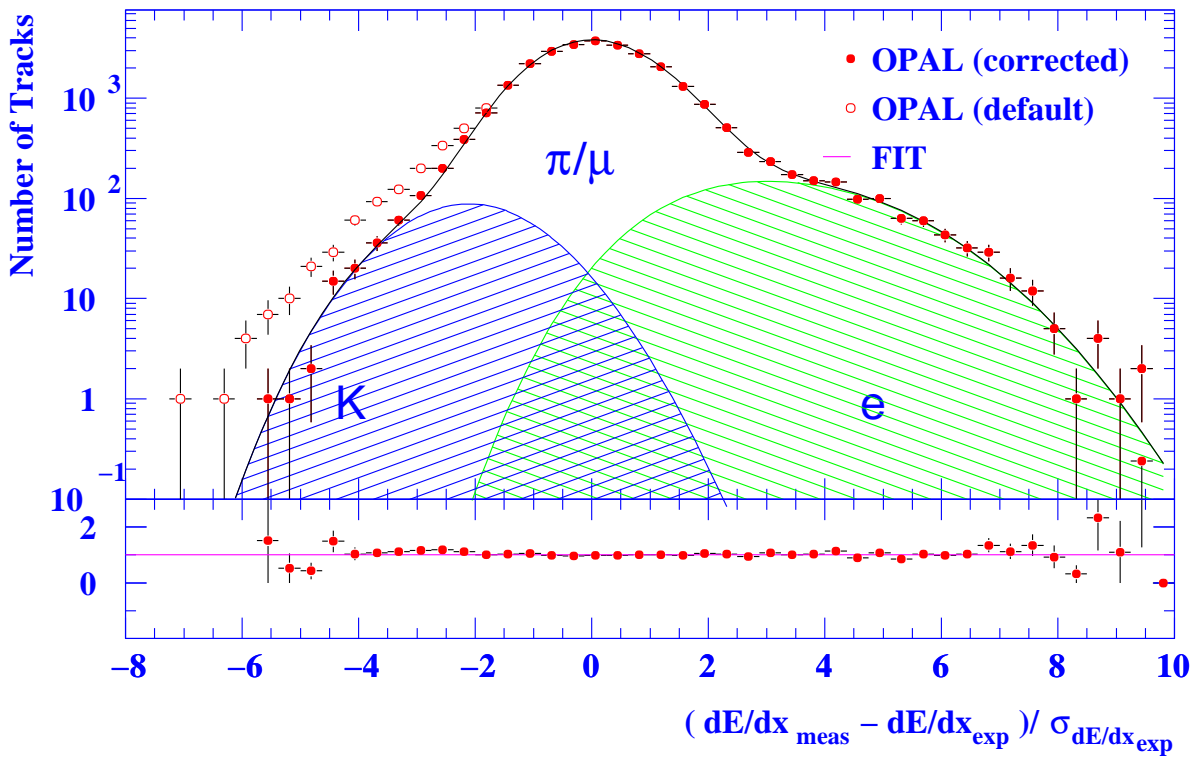}
    \caption{\em Pull distribution obtained under a pion hypothesis for all
      tracks in 3-prong \Pgt\ lepton decays with a minimum momentum of
      $3\,\mathrm{GeV}$ and a minimum number of 20 hits in the dE/dx
      measurement. Only the tracks with the same charge as the
      decaying \Pgt\ lepton are shown. The solid points with error bars are
      data after all corrections and the function shows the
      expectation as explained in the text. The open points in the
      range between -8 to -2 show the same distribution but without
      the corrections mentioned in the text. The smaller plot at the
      bottom shows the ratio of the full data points to the
      sum of the functions.}   
    \label{fig:dE/dxpull}
  \end{center}
\end{figure}

The effect of all corrections can be seen in 
Figure \ref{fig:dE/dxpull}. It shows the {\dedx} pull distribution
under a pion-hypothesis for all like-sign\footnote{Tracks with the
  same charge as the initial \Pgt\ lepton.} 
tracks from 3-prong tau decays. The solid
line is obtained by a fit of three Gaussians to the measured pull
distribution in 9 variable momentum bins from $3\,\mathrm{GeV}$
to $40\,\mathrm{GeV}$. Since the number of muons in this sample is
very small and their energy loss is very similar to the one for pions
in the momentum range considered here, separate Gaussians for pions
and muons are not needed. The free parameters in this fit are two of
the three normalizations (the third one is constrained so that the sum
of all three is equal to the observed number of tracks), one mean to
allow for an overall shift (the position of the other two is
calculated using the corresponding prediction from the Bethe-Bloch
formula) and the width of the Gaussian which describes the
contribution from pions. The deviation of this parameter from unity is
used to obtain correction factors for the error of the energy loss
measurement. The width of the two other Gaussians is calculated
assuming that the relative error is constant.

From the measured energy loss, its error and the expectation
calculated using the Bethe-Bloch equation, $\chi^2$ probabilities
are calculated that the measured energy deposition is in accordance
with the expectation for a given particle type. Pion- and
kaon-weights, $W_\Pgp$ and $W_\PK$, as used in this paper, are then
calculated by taking one minus the value of this probability. These
weights acquire a sign depending on whether the actual energy loss
lies above or below the expectation for a certain particle
hypothesis. This means that $W_\Pgp$ is expected to be close to $-1$
for kaons since their energy loss per unit length is smaller in the
momentum range relevant in this analysis. For electron tracks,
$W_\Pgp$ is expected to be close to $+1$ due to the higher energy loss
in this case. Whenever these quantities are used in the selection, a
cut on at least $20$ {\dedx} hits for this track is made implicitly. 

\subsection{Photon Reconstruction and Identification of Neutral Pions}
\label{sub:Npi0} 
\begin{figure}[!b]
  \begin{center}
    \includegraphics[width=.80\textwidth]{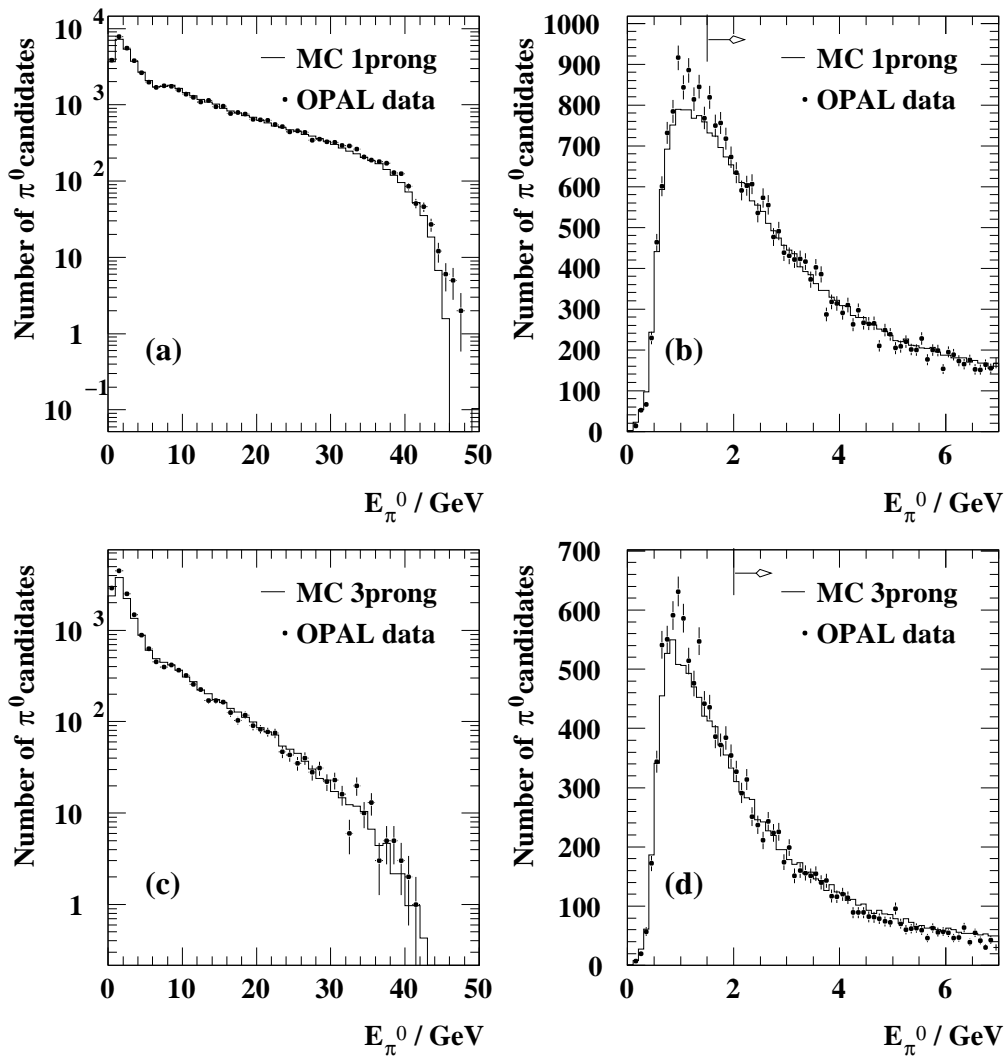}
    \caption[Energy Distribution of the \Pgpz\ Selection]
            {\em Energy distribution of the reconstructed
            \Pgpz\ candidates. Plots (a) and (b) show the
            reconstructed energy for events with one track,
            plots (c) and (d) for events with three tracks. All plots
            are normalized to the number of \Pgt\ lepton
            decays. The arrows indicate the region selected.} 
    \label{fig:Eg-13prong}
  \end{center}
\end{figure}

The reconstruction of \Pgpz\ mesons from photon candidates starts from
an algorithm that has been used in previous OPAL publications (see e.g.
\cite{bib:OPALSven}).  It is based on the analysis of shower
profiles in the ECAL as a function of the energy and direction of
photons. In the fit each cluster is treated individually. Since the
lead glass blocks in the barrel have a quasi-pointing\footnote{The
  longitudinal axis is pointing towards the interaction region. The
  blocks are tilted slightly from a perfectly pointing geometry to
  prevent neutral particles from escaping through the gaps between
  blocks.} geometry, only the lateral shower profile is
considered. This algorithm provides the number of photon candidates
and their corresponding 3-vectors that best describe the measured
profile of the electromagnetic cluster. All photon candidates are
assumed to originate from the interaction point (approximated by the
primary vertex).   
Since the relative branching fractions of decays with and without
\Pgpz\ mesons are different in strange and in non-strange final states, the
parameters of the algorithm were optimized for this analysis as described
below. 

Within each cluster only blocks with an energy of at least
$150\,\mathrm{MeV}$ are considered and the total cluster energy must
exceed $600\,\mathrm{MeV}$. The expected mean energy deposited by a
minimum ionizing particle is then subtracted from all ECAL blocks hit
by a charged particle before the fit.  
If a photon candidate is too close to the entrance point of a track
into the ECAL, the track's  hadronic interaction can distort the
photon energy measurement. Therefore, a minimum angle of $2.8^\circ$
between a photon candidate and a track's ECAL entrance point is required.
This value is obtained from studies of the rate of fake \Pgpz\ mesons in the
decay {$\tauK$} and subsequent optimization. 
To improve the energy resolution and the purity of the selection, a
pairing algorithm is applied to recombine fake photon candidates with
the closest photon candidates that were wrongly split up by the
reconstruction procedure. The recombination is performed using the 
Jade jet finding scheme with the P0 option \cite{bib:Bethke}.
Jet resolution parameter $y_\mathrm{cut}$ values are optimized to
obtain the best description of the number of expected photons using
Monte Carlo events. Two $y_\mathrm{cut}$ values are determined, one
for clusters where a track is pointing to one of the blocks in the
cluster and one for those without tracks. This is necessary since the
hadronic interaction of charged particles disturbs the shower
profile. The optimized $y_\mathrm{cut}$ values are $-3$ and $-4.6$ for
clusters with and without tracks, respectively. The angular resolution
of this algorithm is $2.1^{\circ}$ for clusters with tracks and
$1.7^{\circ}$ for clusters without tracks. Since these opening angles
correspond directly to the energy of the neutral pion, photon candidates with
an energy of more than $7.5\,\mathrm{GeV}$ are directly interpreted as
neutral pion and the 4-vector is corrected to account for the \Pgpz\
mass. The energy of the reconstructed \Pgpz\ candidates for events
with one and three tracks is shown in Figure \ref{fig:Eg-13prong}. The
plots are normalized to the number of \Pgt\ decays in the event
sample.  
Neutral pion candidates with an energy below
$1.5\,\mathrm{GeV}$ in the 1-prong case and $2\,\mathrm{GeV}$ in the
3-prong case are rejected.  

For the remaining photon candidates, all two-photon combinations are
tested. The combination which results in the maximum number of neutral
pions with invariant two-photon masses not exceeding the \Pgpz\ mass by
more than $1.5\sigma$ is retained. A fit with a \Pgpz\  mass
constraint is then applied to all \Pgpz\ candidates. All \Pgpz\
candidates are assumed to originate from the primary interaction
point. 

\subsection{Identification of \PKzS\ }\label{sub:IKOs}
\enlargethispage*{.5cm}
\begin{figure}[!t]
  \begin{center}
    \includegraphics[width=.95\textwidth]{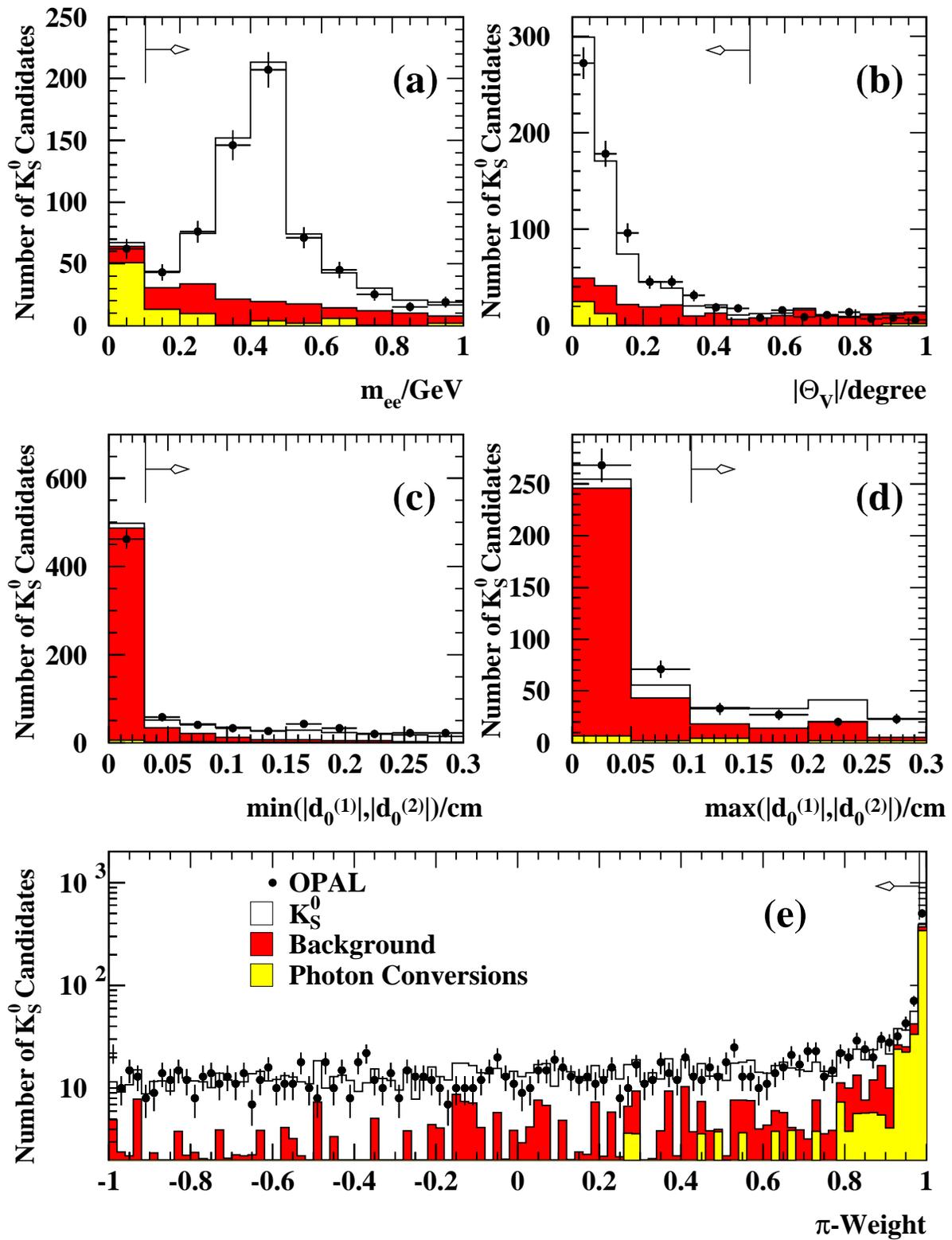}
    \caption[Variables used in the \PKzS\ selection]
            {\em Variables used in the \PKzS\ selection. A detailed
            description of all variables is given in the text. 
            The dots represent the data and the open histogram is
            Monte Carlo signal. The shaded areas show the background where
            photon conversions are marked separately. The arrows
            indicate the region selected. For all plots, all selection
            cuts have been applied except for the cut on the variable
            shown. All plots are normalized to the number of \Pgt\
            decays.} 
    \label{fig:K0_1}
  \end{center}
\end{figure}
\begin{figure}[!t]
  \begin{center}
    \includegraphics[width=.8\textwidth]{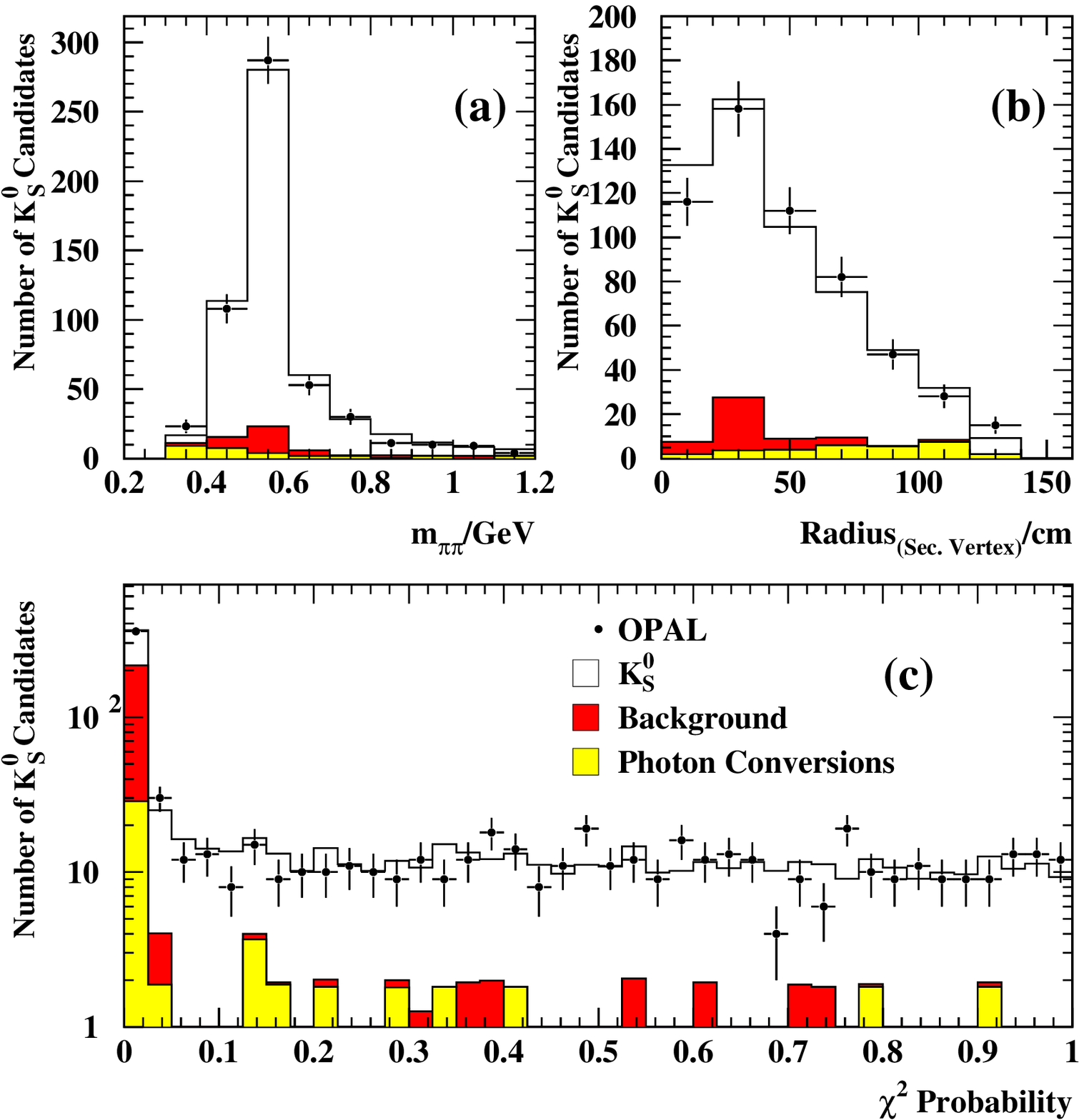}
    \caption[Result the \PKzS\ selection]
            {\em Result of the \PKzS\ selection. Plot (a) shows the
            invariant mass distribution of the \PKzS\ candidates under pion
            hypothesis before the kinematic fit. 
            Plot (b) shows the
            radius of the reconstructed secondary vertex and plot (c)
            the distribution of the $\chi^2$-probability of the 2C-fit. A cut
            is applied on the probability at $10^{-5}$. The dots represent the
            data and the open histogram is the Monte Carlo signal. The
            shaded areas show the background where photon conversions are
            marked separately. All plots are normalized to the number
            of \Pgt\ decays.}  
    \label{fig:K0_2}
  \end{center}
\end{figure}

The \PKz\ signal consists of 50\% \PKzL\ and 50\% \PKzS. The signature of a
\PKzL\ decay is a large energy deposit in the hadron calorimeter without
an associated track pointing to the cluster. The resolution of the OPAL hadron
calorimeter would not allow for a clean reconstruction of this channel, thus it
is not considered here. For the \PKzS, two decay modes are dominant,
{$\KOsppn$} ($\approx 32\%$) and {$\KOspp$} ($\approx 68\%$). In this
analysis only the latter is considered since the photon reconstruction
algorithm exploits the quasi-pointing geometry of the electromagnetic
calorimeter in the barrel (see Section \ref{sub:Npi0}). Thus only photons from
the primary vertex can be properly reconstructed.  

The selection starts by combining each pair of oppositely charged
tracks. Each track must have a transverse momentum with respect to the
beam axis of $p_\mathrm{T}\ge 150\,\mathrm{MeV}$, a minimum of 20 out
of 159 possible hits in CJ, at least $20\%$ of all geometrically
possible hits and a maximum $\chi^2$ for the track fit of 50. For each
combination of tracks, their intersection points in the plane
perpendicular to the beam axis are calculated. The one with a radius
less than $150\,\mathrm{cm}$ is selected as the secondary vertex. If
two vertices are found that satisfy this condition, the one with
the first measured hit closest to the intersection point is
selected. In addition, the $z$-coordinate of the vertex has to satisfy
$|z_\mathrm{V}|<80\,\mathrm{cm}$.  

The variables used in the selection described below are shown in
Figure \ref{fig:K0_1}. For each \PKzS\ candidate, the angle between the
reconstructed momentum of the candidate vertex and the \PKzS\ 
direction of flight must be less than
$|\Theta_\mathrm{V}|<0.5^\circ$. If the number of {\dedx}-hits is more
than 20 on at least one of the tracks, the energy loss information is
used to veto photon conversions. All candidates where at least one
track has a pion weight of more than $0.98$ are rejected. To further
reduce the background from photon conversions, the invariant mass of
the \PKzS\ candidate is calculated using electron hypothesis for both
tracks. All combinations with a mass $m_{\Pe\Pe}<0.1\,\mathrm{GeV}$
are rejected. The remaining background that mainly comes from 3-prong
\Pgt\ lepton decays or from wrong track combinations is rejected by
applying cuts on the impact parameters (with respect to the nominal
interaction point $d_0$) of the two tracks. The $d_0$ values must have
opposite sign according to the OPAL convention and the absolute values
have to satisfy $\min(|d_0^{(1)}|,|d_0^{(2)}|)>0.03\,\mathrm{cm}$ and 
$\max(|d_0^{(1)}|,|d_0^{(2)}|)>0.1\,\mathrm{cm}$.   

The remaining \PKzS\ candidates must have a momentum of
$p_\PKzS >3\,\mathrm{GeV}$. A 3D vertex fit is applied to each
candidate, that includes a constraint of the invariant two-track mass
under the pion hypothesis to the nominal \PKzS\ mass.
This is a 2C fit and a cut on the $\chi^2$ probability at $10^{-5}$ is
applied. The invariant two-track mass under the pion hypothesis before the
kinematic fit can be found in Figure \ref{fig:K0_2} together with the
radius of the reconstructed secondary vertex and the $\chi^2$
probability of the kinematic fit. In Figure \ref{fig:K0_2}(a) the
$\pi\pi$-invariant mass spectrum is shown without any vertex
constraint. If more than one \PKzS\ candidate shares the same track,
the one with the smallest deviation from the nominal \PKzS\ mass before
the fit is selected. 

After this selection procedure, a total of 535 \PKzS\ candidates
remain with an estimated purity of 82\%. About $70\%$ of the
background consists of wrong combinations of tracks, and $30\%$ comes from
photon conversions. In one data event, two \PKzS\ candidates are found
within one cone. This event is considered to be background. 
\section{Identification of \Pgt\ Final States}\label{sec:SelEv}
For the selection of the various final states, a cut-based procedure is
used where each \Pgt\ decay is treated independently. For all selected
decay modes, the cone axis, calculated from the momenta of all tracks
and neutral clusters identified in the electromagnetic calorimeter,
must have a polar angle within $|\cos\theta|<0.68$ for the reasons
explained above. Each selected cone must have at least one good
track coming from the interaction point and the summed momenta of all
tracks have to be less than the beam energy. Since there is at least
one hadron in the final states considered here, the total energy
deposited in the hadron calorimeter within the cone is required to
exceed $1\,\mathrm{GeV}$. 

\subsection{Corrections to the Invariant Mass Spectra}\label{sub:MCorr}
The spectral function from \Pgt\ lepton decays is a weighted invariant
mass distribution of all hadronic final states with strangeness. The
various final states have different experimental resolutions and
migration effects that must be corrected for.
To correct the observed data, the following method was used. The
elements $c_{ij}$ of the inverse detector response matrix were
determined directly from Monte Carlo. They represent the probability
that an event reconstructed in bin $j$ was generated in bin $i$. To
calculate this probability  Monte Carlo samples for the signal
channels with mass distributions according to phase space were used. 
The corrected distribution was then obtained by   
\begin{equation}
  g_i^\mathrm{Corrected} = \sum_j c_{ij} (g_j^\mathrm{DATA} -
  g_j^\mathrm{Background})
\end{equation}
where $g_j^\mathrm{DATA}$ is the number of events in the data with a
reconstructed mass in bin $j$, $g_j^\mathrm{Background}$ is the number
of background events predicted in bin $j$ and $g_i^\mathrm{Corrected}$
the number of events in mass bin $i$ after correction. 

The corrected distributions will in general be biased towards the
Monte Carlo input distributions. To reduce the bias from this
approach, the method was applied iteratively. The result of the
preceding iteration was used to refine the elements of the inverse
detector matrix. The optimal number of iterations was determined, using
Monte Carlo simulations, to be two for all final states considered
here. The corresponding systematic uncertainties are discussed  in
Section \ref{sub:Syst}. Finally, detection efficiency corrections were
applied to the corrected invariant mass distributions.
The bin width of $150\,\mathrm{MeV}$ of the measured invariant mass
spectra were chosen according to the mass resolution of the \Pgt\
decay channels measured.

\subsection{$\Kpm$ Final States}\label{sub:Kp}
The $\Kpm$ mass spectrum consists of two measured modes, {$\Kpn$} and
{$\KOp$}. From the latter decay mode, only the decays {$\KOspp$} are
measured. The \PKst\ dominance is well established in both cases.  

\subsubsection{$\Kpn$}
\begin{figure}[!b]
  \begin{center}
    \includegraphics[width=.75\textwidth]{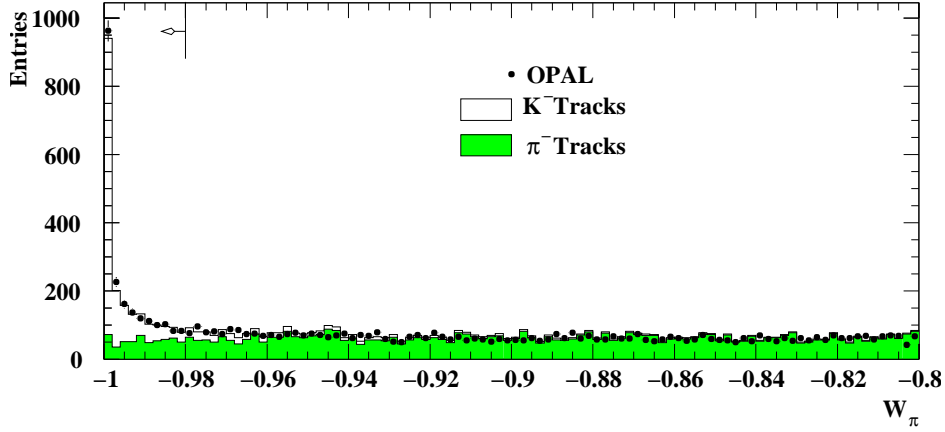}
    \caption[Variables used in the {$\KOp$} selection]
    {\em The pion weight $W_\pi$ for 1-prong \Pgt\
      decays in the range $(-1:-0.8)$. The dots are the data
      points. The open (shaded) histogram denotes the contribution from
      kaon (pion) tracks as predicted by the Monte Carlo
      simulation. The plot is normalized to the number of \Pgt\
      decays. Events on the side of the direction of the arrow are
      considered kaon candidates.}
    \label{fig:wp1prong}
  \end{center}
\end{figure}

\begin{figure}[!b]
  \begin{center}
    \includegraphics[width=.75\textwidth]{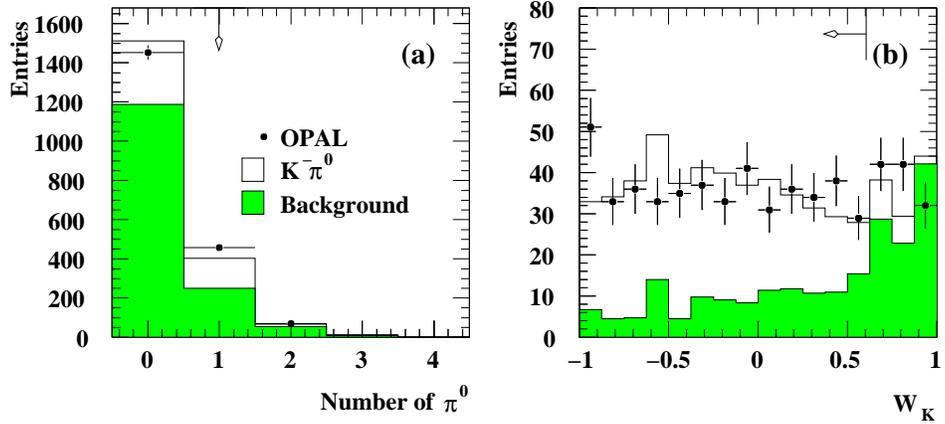}
    \caption[Variables used in the {$\Kpn$} selection]
    {\em Variables used in the {$\Kpn$} selection. The dots
      are the data points and the open histogram is the
      prediction from the  Monte Carlo.
      Plot (a) shows the number of reconstructed \Pgpz\ mesons with
      $E>1.5\,\mathrm{GeV}$. In this plot the shaded area is the
      background prediction from the Monte Carlo. Plot (b)
      shows the kaon weight as explained in the text. Here the
      shaded area represents the expected background from pion
      tracks.  The arrows indicate the events kept in the
      selection. For both plots, all selection cuts have been
      applied except for the cut on the variable shown.}
    \label{fig:kpi0_1}
  \end{center}
\end{figure}

In the {$\Kpn$} selection, exactly one good track coming from the
primary vertex is required. This track must have a minimum
momentum of $p>3\,\mathrm{GeV}$. For the track to be selected as a
kaon, the pion weight has to satisfy $W_\pi< -0.98$ and the kaon
weight $W_\PK<0.6$. Furthermore, exactly one identified neutral pion
is required with $E_\mathrm{min}^\Pgpz >1.5\,\mathrm{GeV}$ selected
using the algorithm explained in Section \ref{sub:Npi0}. The variables
used in this selection are shown in Figures \ref{fig:kpi0_1} and
\ref{fig:wp1prong}.

\begin{figure}[!t]
  \begin{center}
    \includegraphics[width=.75\textwidth]{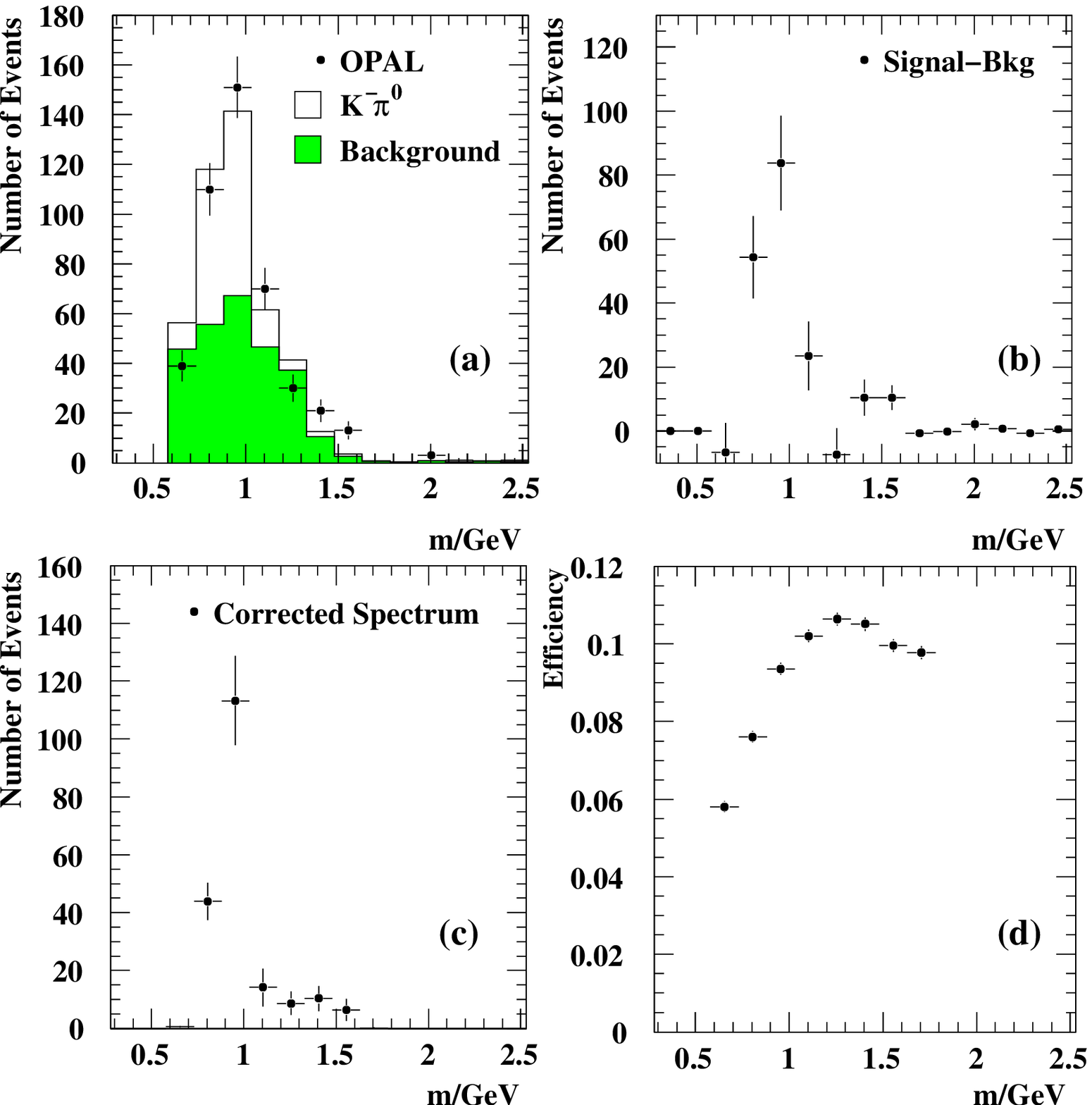}
    \caption[Variables used in the {$\Kpn$} selection]
    {\em Result of the {$\Kpn$} selection. Plot (a) shows
      the measured invariant mass spectrum. The dots are the data, the
      open histogram is the Monte Carlo signal and the shaded area is the
      background. Plot (b) shows the background subtracted
      spectrum, (c) shows the corrected
      spectrum. Note that the correlation between neighboring bins
      in the corrected spectrum is of the order of $50\%$. Plot (d) is
      the selection efficiency as a function of the invariant mass.} 
    \label{fig:kpi0_2}
  \end{center}
\end{figure}

From this selection $360$ events are seen in the data with $190.5$
background events predicted from Monte Carlo. This corresponds to a
background fraction of $54\%$. The invariant mass spectrum can be
seen in Figure  \ref{fig:kpi0_2}. The mass resolution in this
channel is approximately $40\,\mathrm{MeV}$. The main background
comes from $\tauK$ decays where one fake neutral pion was
reconstructed. Additional sources of background are {$\tauppn$},
{$\tauKKO$} and {$\tauKOKpn$} where the \PKz\ is a \PKzS\ decaying
to two neutral pions or a \PKzL\ which does not decay within the jet chamber. 

\subsubsection{$\KOp$}
\begin{figure}[!t]
  \begin{center}
    \includegraphics[width=.75\textwidth]{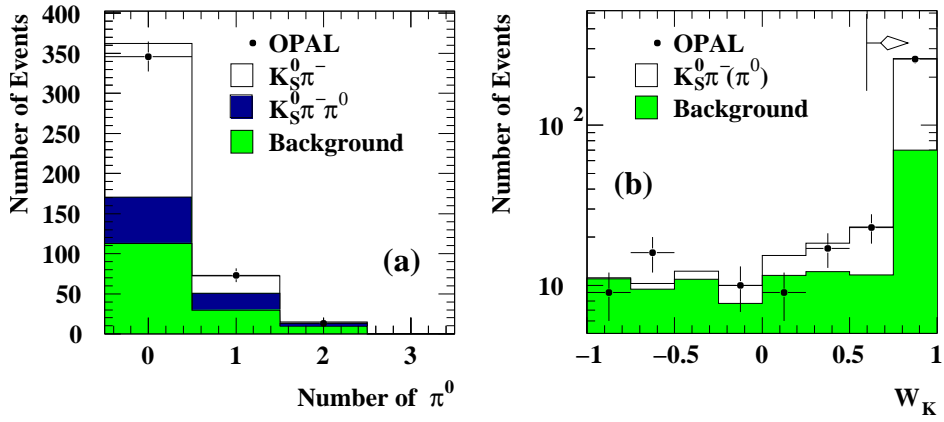}
    \caption[Variables used in the {$\KOp$} selection]
    {\em Variables used in the {$\KOp$} and {$\KOppn$}
      selection. (a) shows the number of reconstructed
      \Pgpz\ mesons. The dots are the data, the open histogram is the
      signal. The dark-shaded area represents the contribution from
      \KOppn\ final states, the light-shaded area denotes other
      background channels. (b) is the kaon weight of the primary
      track}
    \label{fig:k0pi_1}
  \end{center}
\end{figure}

The selection is very similar to that for the {$\Kpn$} final state. Here
exactly one identified \PKzS\ is required using the procedure from Section
\ref{sub:IKOs}. In addition, one good track from the primary
vertex is required. If the momentum of this track lies above the
kinematically allowed minimum for a kaon, the same identification
procedure as mentioned above is applied to veto decays
{$\tauKKO$}. Only events with zero reconstructed \Pgpz\ mesons are
accepted. The variables used in the selection are shown in Figure
\ref{fig:k0pi_1}.   

\begin{figure}[!b]
  \begin{center}
    \includegraphics[width=.75\textwidth]{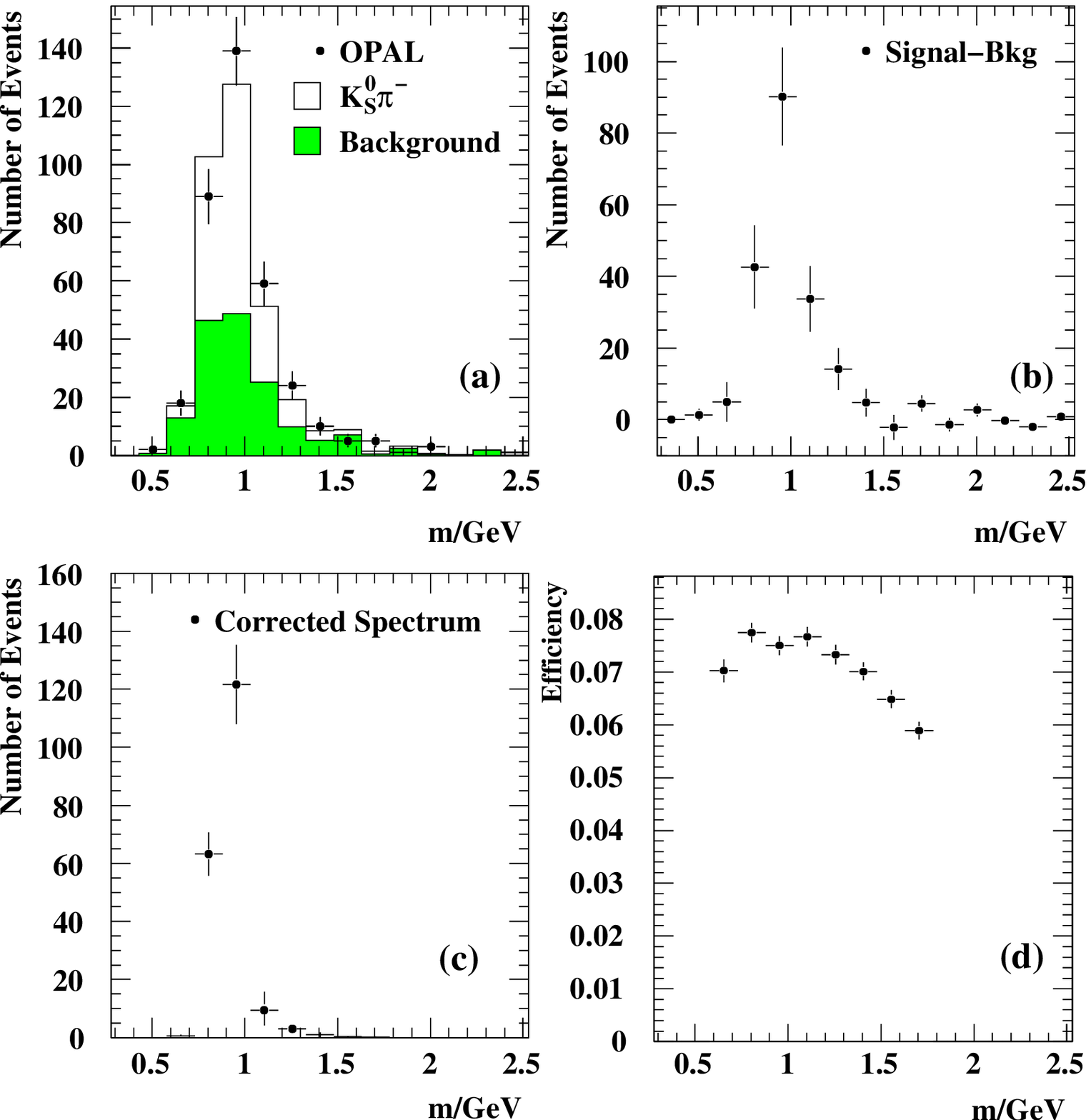}
    \caption[Variables used in the {$\Kpn$} selection]
    {\em Result of the {$\KOp$} selection. Plot (a) shows
      the measured invariant mass spectrum. The dots are the data, the
      open histogram is the Monte Carlo signal and the shaded
      area is the background. Plot (b) shows the background
      subtracted spectrum, (c) the corrected spectrum. 
      Note that the correlation between neighboring bins in the
      corrected spectrum is of the order of $50\%$. Plot (d)
      is the selection efficiency as a function of the invariant
      mass.}  
    \label{fig:k0pi_2}
  \end{center}
\end{figure}

From this selection $361$ events are expected
with a background fraction of $47\%$, and $344$ are seen in the
data. The main background contributions come from decays
{$\tauKOKOp$}, {$\tauKKO$} and {$\tauKOppn$} where the neutral pion
escapes detection. The invariant mass spectrum for this channel is
shown in Figure \ref{fig:k0pi_2}. The mass resolution in this
channel is approximately $60\,\mathrm{MeV}$.  

\clearpage 

\subsection{$\Kppm$ Final States}\label{sub:Kpp}
The $\Kppm$ final state consists of the decay modes: {$\Kpp$}, {$\KOppn$}
and {$\Kpnpn$}. To select these final states, the following procedure
is applied. 

\subsubsection{$\Kpp$}
The selection starts by requiring exactly three good tracks coming from the
interaction point. These tracks are fitted to a common vertex and the
fit probability is required to be larger than $10^{-7}$. In
addition, each pair of oppositely charged tracks has to fail the 
selection criteria for neutral kaons as defined in Section
\ref{sub:IKOs}. These two requirements reduce the background from
photon conversions and decays containing \PKzS.   

\begin{figure}[!b]
  \begin{center}
    \includegraphics[width=.8\textwidth]{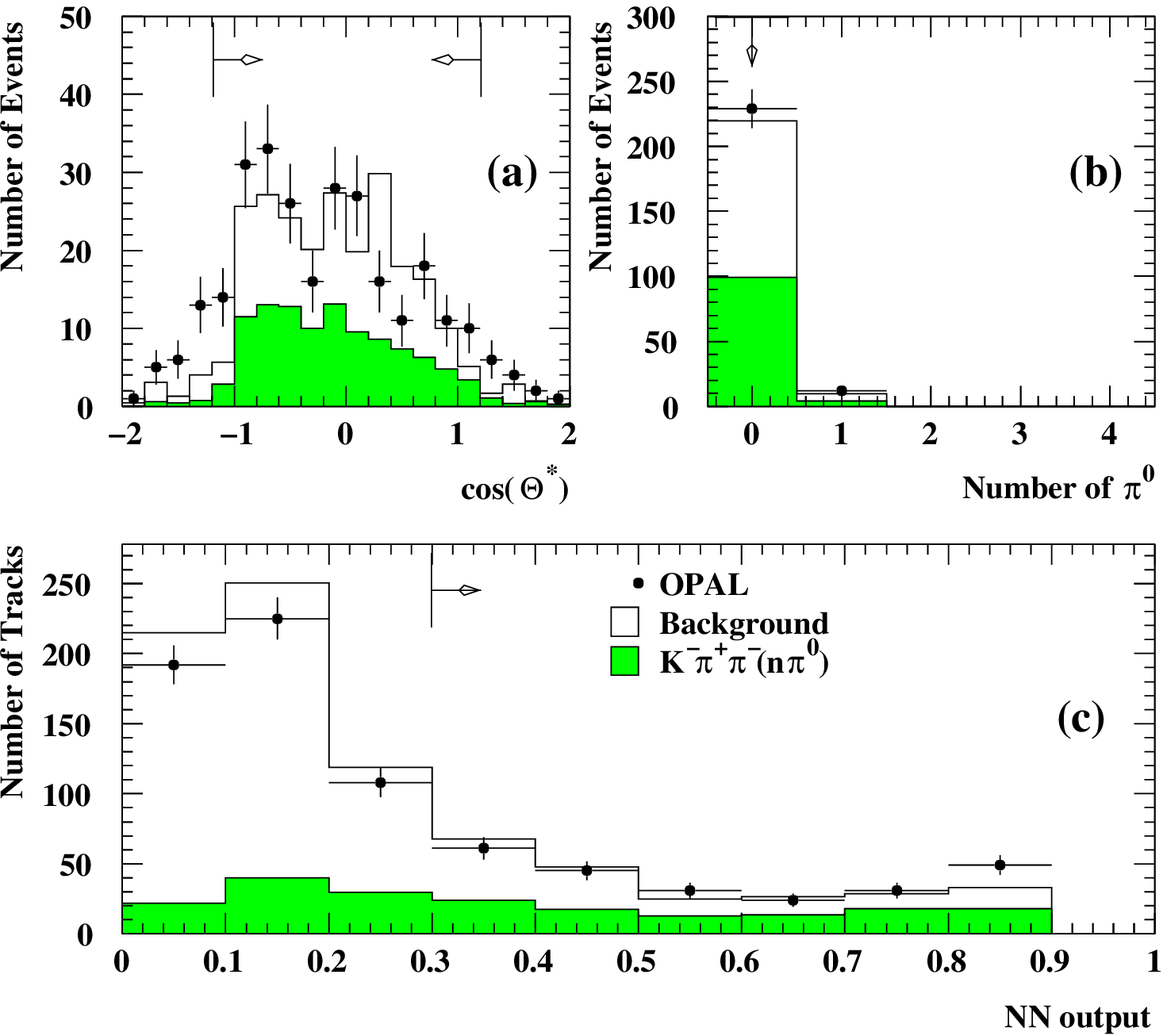}
    \includegraphics[width=.78\textwidth]{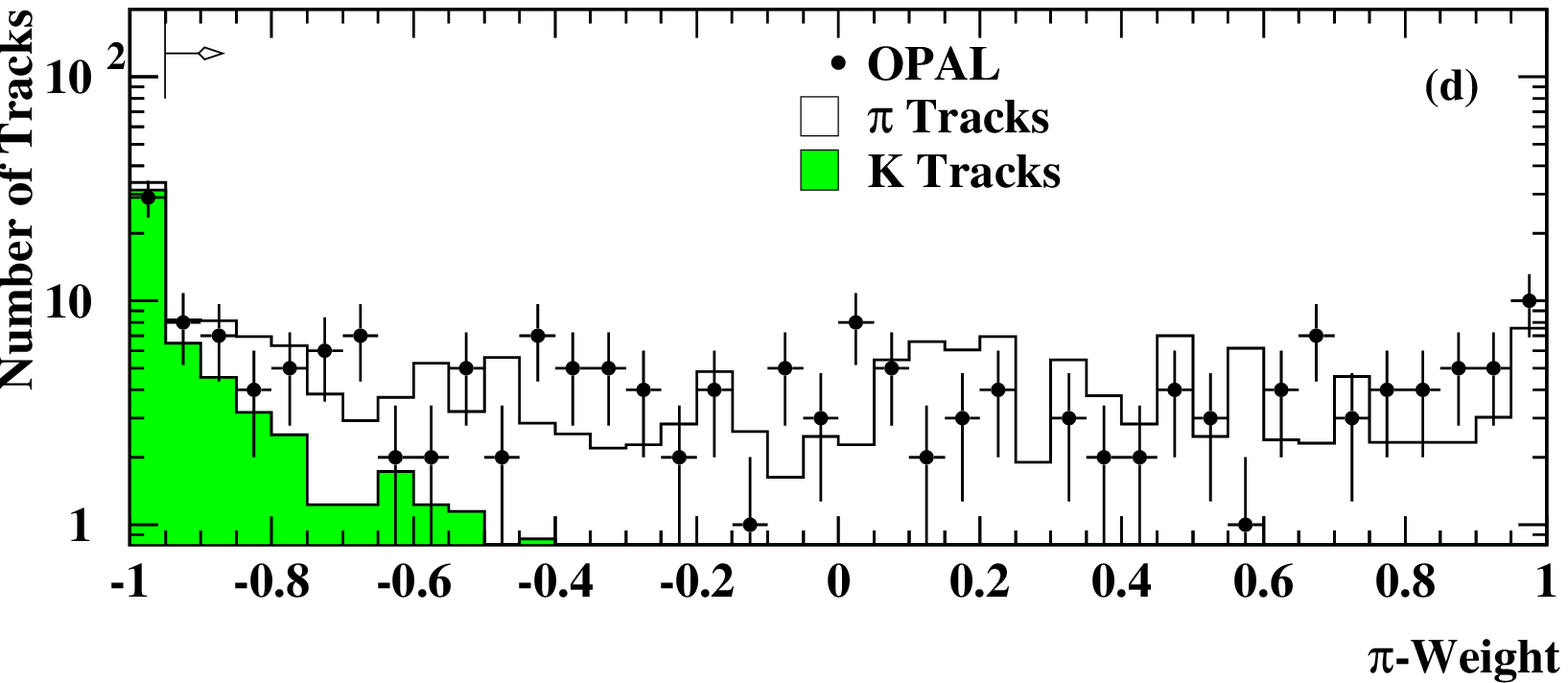}
    \caption[Variables used in the {$\Kpp$} selection]
    {\em Variables used in the {$\Kpp$}
      selection. Plot (a) shows the cosine of the 
      Gottfried-Jackson angle $\Theta^\ast$, (b) the number of 
      reconstructed neutral pions and (c) the output of the
      neural network. The dots are the data, the open
      histogram is the background prediction from the Monte Carlo and
      the shaded histogram denotes the contribution from \KppNpn\
      final states. (d) shows the pion weight of the unlike-sign
      track in the $\Kpp$ selection. Here the open histogram denotes
      the contribution from pion tracks, the shaded histogram is the
      contribution from kaon tracks. The dots are the data.
      For all plots, all selection cuts have been applied except for
      the cut on the variable shown. The arrows indicate the region
      selected.}
    \label{fig:kpipi_1}
  \end{center}
\end{figure}

To identify the kaon in the decay, one of the like-sign candidate
tracks must have $p>3\,\mathrm{GeV}$ and $W_\pi<-0.9$. To further
reduce the pion background among these candidate tracks, $W_\PK$ and
$W_\pi$ are input to a neural network. The track is rejected if the
output of the neural network is below $0.3$ (see Figure
\ref{fig:kpipi_1}(c)). Exactly one like-sign track is allowed to fulfill
these requirements, otherwise the decay is treated as background.  
If the momentum of the unlike-sign track is consistent with the {$\tauKKp$}
hypothesis, this \Pgt\ decay is only accepted if $W_\pi>-0.95$ (see
Figure \ref{fig:kpipi_1}(d)). 

The algorithm for identifying neutral pions (Section \ref{sub:Npi0})
is then applied to the selected cones. For this channel, the number
of reconstructed \Pgpz\ mesons with an energy greater than
$2\,\mathrm{GeV}$ is required to be zero (see Figure
\ref{fig:kpipi_1}(b)). Otherwise this \Pgt\ decay is treated as
background. 
To further improve the purity of the selection, the cosine of the
decay angle in the rest frame of the \Pgt\ lepton, the so-called
Gottfried-Jackson angle $\Theta^\ast$ is calculated. The
$\cos\Theta^\ast$ distribution is shown in
Figure \ref{fig:kpipi_1}(a). For events where the kaon hypothesis was
applied to the wrong track or the number of identified neutral pions does
not correspond to the true number, this calculation leads to
unphysical values of that variable. Due to resolution effects,
correctly identified signal events can also give values beyond $\pm
1$. Therefore, a cut was applied at $\cos\Theta^\ast=\pm 1.2$. 
The contribution from $\tauKppNpn$ events is included in the
background estimate. 

\begin{figure}[!h]
  \begin{center}
    \includegraphics[width=.85\textwidth]{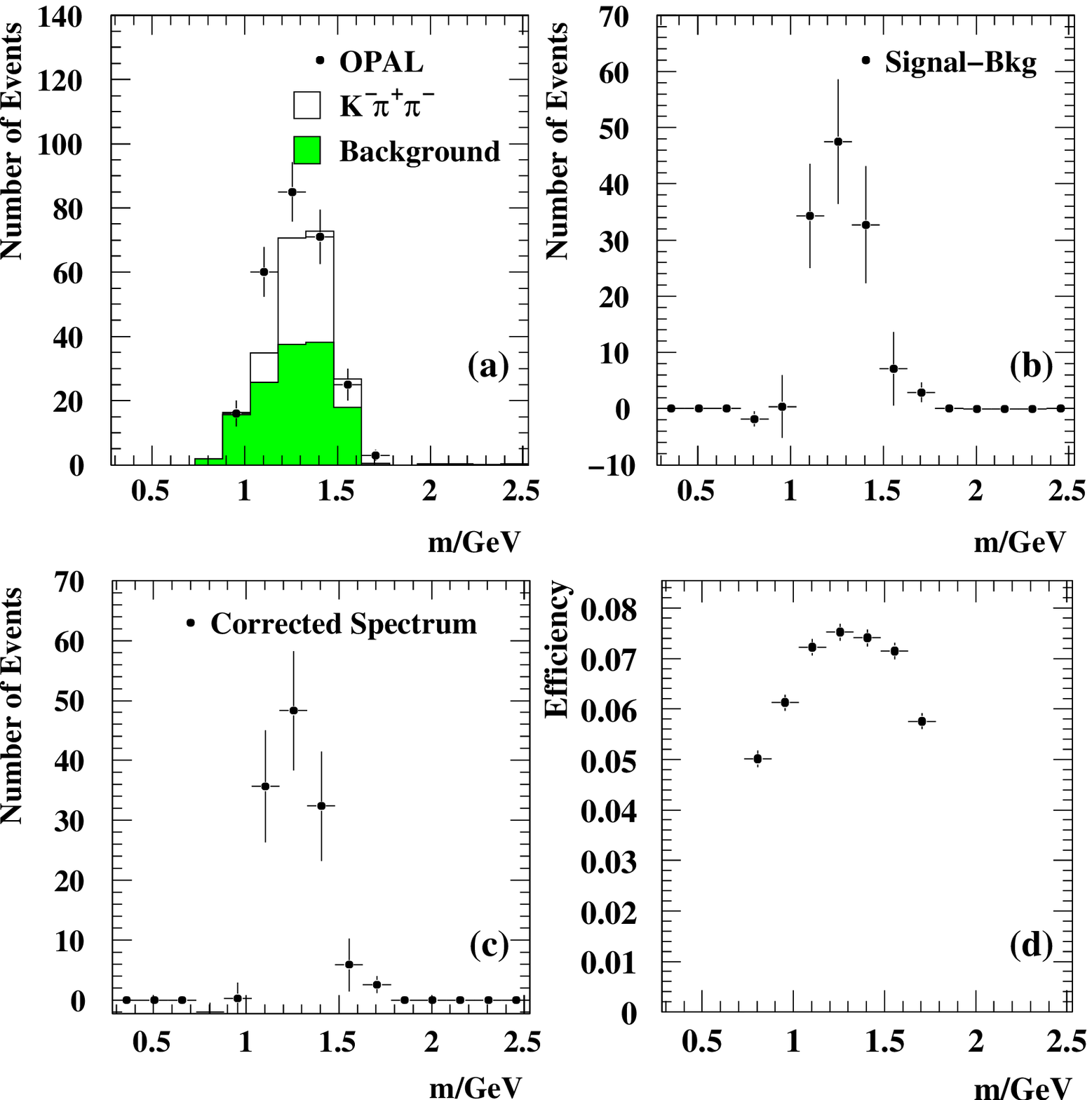}
    \caption[Variables used in the {\Kpn} selection]
    {\em Result of the {\Kpp} selection. Plot (a) shows
      the measured invariant mass spectrum. The dots are the
      data, the open histogram is the Monte Carlo signal and
      the shaded area is the background. Plot (b) shows the
      background subtracted spectrum, plot (c) is the corrected
      spectrum. Note that the correlation between neighboring bins
      in the corrected spectrum is of the order of $50\%$. Plot (d)
      is the selection efficiency as a function of the invariant
      mass.}   
    \label{fig:kpipi_2}
  \end{center}
\end{figure}

From this selection $269$ events are seen in the data with a
contribution of $149.8$ background events predicted from Monte
Carlo. This corresponds to a background fraction of $63\%$. The main
background contribution comes from decays {$\tauppp$}, {$\tauKKp$}
and {$\tauKpppn$}, where the \Pgpz\ meson escapes detection. The invariant
mass spectrum can be found in Figure \ref{fig:kpipi_2}. The mass
resolution in this channel is approximately $20\,\mathrm{MeV}$. 

\subsubsection{\KOppn}
Exactly one identified \PKzS\ and exactly one \Pgpz\ meson with an
energy $E^\Pgpz>2\,\mathrm{GeV}$ is required for this final
state. The pion candidate track has to satisfy the same requirements
as for $\Kpm$ final states.  

\begin{figure}[!hb]
  \begin{center}
    \includegraphics[width=.85\textwidth]{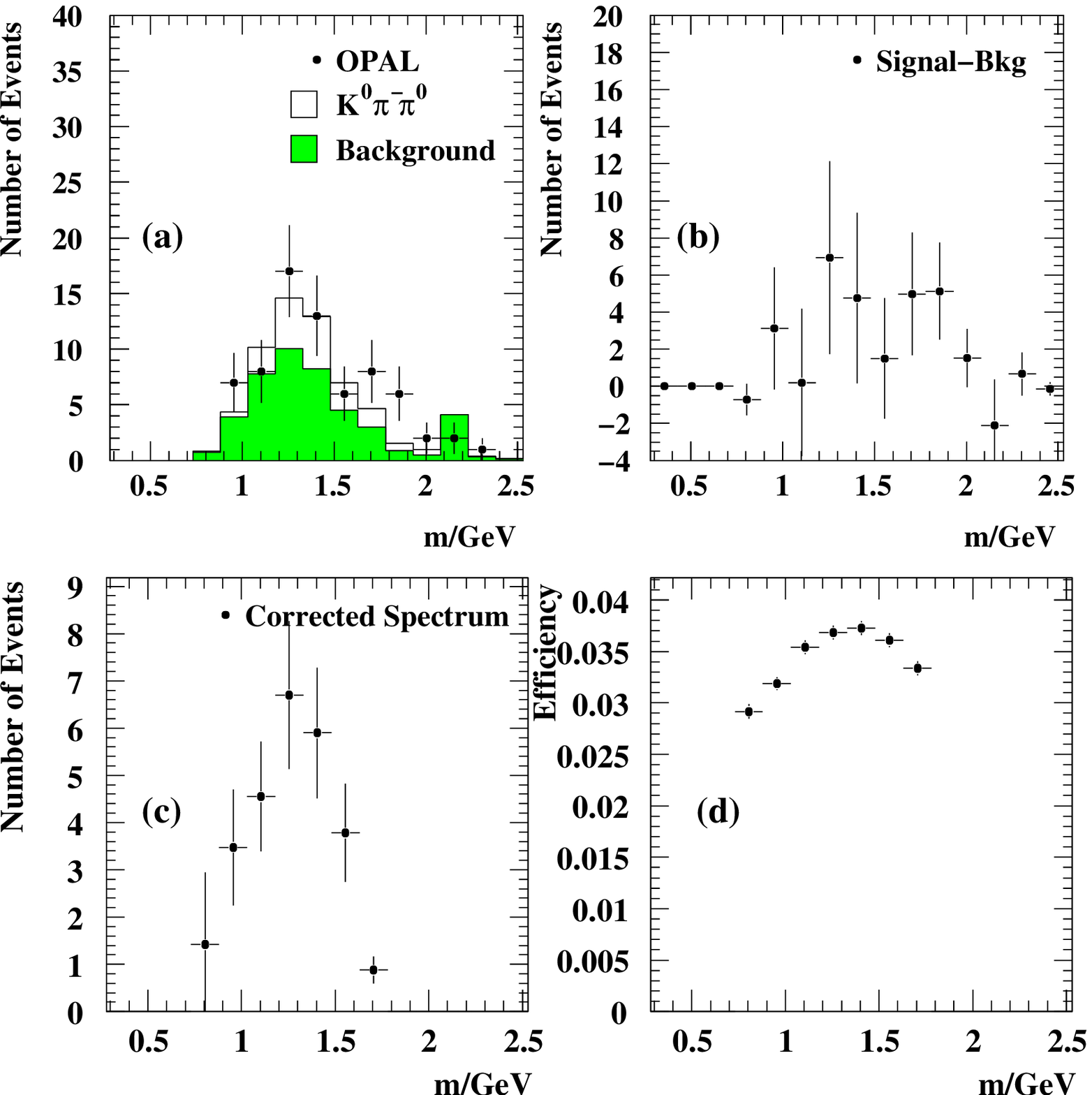}
    \caption[Variables used in the {\Kpn} selection]
    {\em Result of the {\KOppn} selection. Plot (a) shows
      the measured invariant mass spectrum. The dots are the
      data, the open histogram is the Monte Carlo signal and
      the shaded area is the background. Plot (b) shows the
      background subtracted spectrum, (c) the corrected
      spectrum. Note that the correlation between neighboring bins
      in the corrected spectrum is of the order of $50\%$. Plot (d)
      is the selection efficiency as a function of the invariant mass.}
    \label{fig:k0pipi0_1}
  \end{center}
\end{figure}

From this selection, $65$ events are expected and $67$ are seen in
the data with a background fraction of $72\%$. The main background
contribution comes from decays {$\tauKOp$} where the neutral pion escapes
detection. The invariant mass spectrum can be seen in Figure
\ref{fig:k0pipi0_1}. The mass resolution in this channel is
approximately $100\,\mathrm{MeV}$.  

\clearpage 

\subsection{$\Kpppm$ Final States}\label{sub:Kppp}
The $\Kpppm$ signal consists of the following final states: {$\Kpppn$},
{$\KOppnpn$}, {$\Kpnpnpn$} and {$\KOppp$}. From these, only the first
one which has the highest branching fraction is investigated.

\begin{figure}[!hb]
  \begin{center}
    \includegraphics[width=.85\textwidth]{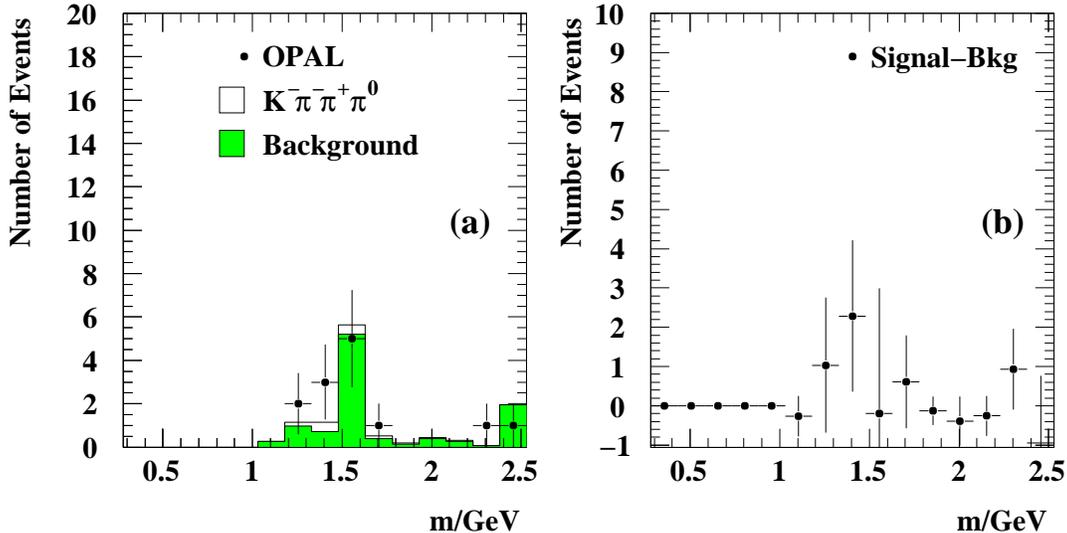}
    \caption[Variables used in the {$\Kpn$} selection]
    {\em Result of the {$\Kpppn$} selection. Plot (a) shows
      the measured invariant mass spectrum. The dots are the data, the
      open histogram is the Monte Carlo signal and the shaded area is the
      background. Plot (b) shows the background subtracted
      spectrum.}
    \label{fig:kpipipi0_1}
  \end{center}
\end{figure}
The same procedure as for the {$\Kpp$} channel is used. In addition,
one identified \Pgpz\ meson with an energy of more than
$2\,\mathrm{GeV}$ is required. The invariant mass spectrum can be seen
in Figure \ref{fig:kpipipi0_1}. The mass resolution in this channel is
approximately $60\,\mathrm{MeV}$.
From this selection, $14$ events are seen in the data with a
 contribution of $10$ events from background. The selection efficiency
 is of the order of $1\%$. The main background contribution comes from
 {$\tauKpp$} decays, where one fake neutral pion was identified. 

Since the number of signal events in this final state is not
significantly different from zero, this channel is not considered
any further in this analysis. For the spectral function, the
Monte Carlo prediction has been used instead. 


\section{Results}\label{sec:Res}
\subsection{Branching Fractions}\label{sub:BF}
The measured data used in the spectral function analysis allows the
determination of competitive branching fractions for the channels
$\tauKpn$ and $\tauKpp$. 
The branching fractions are determined in a simultaneous $\chi^2$
fit, taking all measured final states into account.
The fit function is
\[
N_i = N_i^{\mathrm{non}-\Pgt} + (1-f_\mathrm{bkg}^{\mathrm{non}-\Pgt}) \cdot N^\Pgt
    \sum_j \varepsilon_{ij} B_j F_j^\mathrm{Bias},
\]
where $i$ is the signal channel under consideration and index $j$
runs over all \Pgt\ decay channels. $N_i$ is the number of decays
observed in the data, i.e. the number of \Pgt\ cones passing the cuts.
The number of decays from the non-$\Pgt$ background is denoted by
$N_i^{\mathrm{non}-\Pgt}$, $f_\mathrm{bkg}^{\mathrm{non}-\Pgt}$ is the
fraction of non-\Pgt\ background, $N^\Pgt$ is the number of \Pgt\ events,
$\varepsilon_{ij}$ the efficiency matrix, $F_j^\mathrm{Bias}$ the bias
factor for the respective channel due to \Pgt\ preselection cuts and
$B_j$ are the fitted branching fractions.  
The branching fractions for the two signal channels were allowed to
vary freely, the branching fractions of all other \Pgt\ decay channels
contributing to the background prediction were allowed to vary only
within their PDG errors \cite{bib:PDG}. The selection quantities used
in the fit together with the contributions from the background
channels and their branching fractions used in the Monte Carlo
simulation are shown in Table \ref{tab:BSum}.  
From the fit, the following results are obtained:
\begin{eqnarray*}
  B(\tauKpn) &=& (0.471\pm0.064_\mathrm{stat}\pm0.022_\mathrm{sys})\%\\
  B(\tauKppOpnex) &=&
  (0.415\pm0.059_\mathrm{stat}\pm0.031_\mathrm{sys})\%,
\end{eqnarray*}
where the possibility that the \Pgpp\Pgpm\ pair in the \Kpp\ final
state comes from a \PKzS\ has been excluded. 
For the estimation of the systematic uncertainty, the following
sources are considered. They are summarized in Table
\ref{tab:BSSum}. The total systematic uncertainty is obtained by
adding the individual contributions in quadrature.
\begin{itemize}
\item Energy loss measurement ($\Delta_{\dedx}$):\\
  In the selection, the specific energy loss \dedx\ is used to
  separate pions from kaons. Cuts on corresponding weights are applied
  which are calculated from the pull distribution
  (see Figure \ref{fig:dE/dxpull}).
  A possible shift in this quantity can lead to a systematic
  misidentification of tracks. The pull distribution is therefore
  shifted within the error on its mean and the selection procedure is
  repeated. The difference between the branching fractions obtained
  with and without the shift applied is the systematic uncertainty.
\item Energy scale in \Pgpz\ reconstruction ($\Delta_E$):\\
  The energy resolution can be tested by measuring the invariant 
  two-photon mass from \Pgpz\ decays. A systematic shift in the observed
  mass in the data compared to the detector simulation can be translated
  into a scale factor for the reconstructed photon energies.  Deviations
  of $(0.5 \pm 0.9)\,\mathrm{MeV}$ from the nominal \Pgpz\ mass have
  been observed \cite{bib:OPALSven}, corresponding to a scale factor
  of $1.004 \pm 0.007$. The energies of the reconstructed photons in
  the Monte Carlo samples were therefore varied by $\pm 0.7\,\%$. The
  difference between the branching fractions obtained with and without
  the variation applied is the systematic uncertainty.
\item Momentum scale ($\Delta_p$):\\
  The systematic uncertainty connected with the momentum scale was
  tested using $\PZz\to\Pgmm\Pgmp$ events \cite{bib:OPALSven}. The
  difference in momentum resolution between data and Monte Carlo as a
  function of $\cos\theta$ was studied. To assess the systematic
  uncertainty $\Delta_p$ in hadronic \Pgt\ decays, all particle
  momenta in the Monte Carlo were varied accordingly. The difference
  in the result with and without this variation is quoted as a
  systematic uncertainty.  
\item Monte Carlo statistics ($\Delta_\mathrm{MC}$):\\
  The precision of the background prediction depends on the Monte Carlo
  statistics used in the selection procedure. Therefore, the number of
  background events selected is varied randomly within its statistical
  uncertainty. The observed spread in the branching fraction due to
  this variation is quoted as a systematic uncertainty.
\item Bias factor ($\Delta_{F^\mathrm{Bias}}$):\\
  The bias factors determined from the Monte Carlo are varied
  by their uncertainty and the branching fractions are
  then refitted. The observed spread due to this variation
  contributes to the total systematic uncertainty.
\end{itemize}

\begin{table}[h]
  \centering
  \begin{tabular}{|l|r@{$\ \pm\ $}l|C@{$\ \pm$}C||C@{$\ \pm$}C|}\cline{1-3}
    \multicolumn{3}{|c|}{\both $\tauKpn$} \\\cline{1-3}
    \up No. of Events                       & \multicolumn{2}{c|}{360} & \multicolumn{4}{c}{~} \\
        Selection Efficiency /\%            & 8.42 & 0.17           & \multicolumn{4}{c}{~} \\
        Preselection Bias Factor            & 1.016  & 0.011           & \multicolumn{4}{c}{~} \\
        Non-\Pgt\ Background Fraction       & 0.006  & 0.004           & \multicolumn{4}{c}{~} \\
        \Pgt\ Background Fraction           & 0.540  & 0.027           & \multicolumn{4}{c}{~} \\\cline{2-7}
        \multicolumn{1}{|r|}{\up\ppn}       & \multicolumn{2}{c|}{13.5\%} & 0.051 & 0.005 & 25.41  & 0.14   \\
        \multicolumn{1}{|r|}{\KKOpn}        & \multicolumn{2}{c|}{ 9.9\%} & 6.0   & 0.3   &  0.155 & 0.020  \\
        \multicolumn{1}{|r|}{\K}            & \multicolumn{2}{c|}{ 8.1\%} & 1.25  & 0.06  &  0.686 & 0.023  \\
        \multicolumn{1}{|r|}{\KKO}          & \multicolumn{2}{c|}{ 7.3\%} & 4.5   & 0.2   &  0.154 & 0.016  \\
        \multicolumn{1}{|r|}{\ppnpn}        & \multicolumn{2}{c|}{ 6.2\%} & 0.07  & 0.01  &  9.17  & 0.14   \\
        \multicolumn{1}{|r|}{\Kpnpn}        & \multicolumn{2}{c|}{ 5.0\%} & 9.6   & 0.5   &  0.058 & 0.023  \\
        \multicolumn{1}{|r|}{\Kpnpnpn}      & \multicolumn{2}{c|}{ 2.6\%} & 8.9   & 0.6   &  0.037 & 0.021  \\
        \multicolumn{1}{|r|}{\down\mbox{other}} & \multicolumn{2}{c|}{1.4\%} & \multicolumn{2}{r||}{~}  & \multicolumn{2}{|r|}{~}   \\\hline\hline
        \both ~       & \multicolumn{2}{c|}{Bkg. Fraction    } & \multicolumn{2}{ c||}{Efficiency /\% } 
                                                               & \multicolumn{2}{ c|}{$B^\mathrm{PDG}/\%$} \\\hline
    \multicolumn{7}{c}{~ }\\\cline{1-3}
    \multicolumn{3}{|c|}{\both $\tauKpp$} \\\cline{1-3}
    \up No. of Events                       & \multicolumn{2}{c|}{269} & \multicolumn{4}{c}{~} \\
        Selection Efficiency/\%             & 6.59 & 0.06          & \multicolumn{4}{c}{~} \\
        Preselection Bias Factor            & 0.953  & 0.013           & \multicolumn{4}{c}{~} \\
        Non-\Pgt\ Background Fraction       & 0.007  & 0.006           & \multicolumn{4}{c}{~} \\
        \Pgt\ Background Fraction           & 0.631  & 0.044           & \multicolumn{4}{c}{~} \\\cline{2-7}
        \multicolumn{1}{|r|}{\up\ppp}     & \multicolumn{2}{c|}{21.6\%} & 0.15 & 0.02 & 9.22  & 0.10    \\
        \multicolumn{1}{|r|}{\KKp}        & \multicolumn{2}{c|}{10.3\%} & 3.9  & 0.2  & 0.161 & 0.019   \\
        \multicolumn{1}{|r|}{\ppppn}      & \multicolumn{2}{c|}{ 8.1\%} & 0.5  & 0.1  & 4.24  & 0.10    \\
        \multicolumn{1}{|r|}{\Kpppn}      & \multicolumn{2}{c|}{ 6.7\%} & 2.7  & 0.2  & 0.064 & 0.024   \\
        \multicolumn{1}{|r|}{\down\mbox{other}} & \multicolumn{2}{c|}{ 16.4\%} & \multicolumn{2}{r||}{~} & \multicolumn{2}{|r|}{~} \\\hline\hline
        \both ~         & \multicolumn{2}{c|}{Bkg. Fraction}  & \multicolumn{2}{ c||}{Efficiency /\% } 
                                                              & \multicolumn{2}{ c|}{$B^\mathrm{PDG}/\%$} \\\hline
  \end{tabular}
  \caption{\it Quantities used in the fit for the branching
    fractions. The errors quoted for efficiency, bias factor and
    background fractions are from Monte Carlo statistics only. The last
    column contains the branching fractions for the background
    channels used in the Monte Carlo simulation \cite{bib:PDG}.}
  \label{tab:BSum}
\end{table}

\begin{table}[!b]
  \centering
  \begin{tabular}{|l|c|c|}\cline{2-3}
    \multicolumn{1}{l|}{\both ~} & $\tauKpn$ & $\tauKpp$ \\\hline
    \up   Energy Loss Measurement (\dedx) & 0.012 & 0.019 \\
          Energy Scale                    & 0.010 & 0.011 \\
          Momentum Scale                  & 0.003 & 0.003 \\
          MC Statistics                   & 0.014 & 0.021 \\
    \down Bias Factor ($F^\mathrm{Bias}$) & 0.004 & 0.005 \\\hline\hline
    \both Total                           & 0.022 & 0.031 \\\hline
  \end{tabular}
  \caption{\it Individual contributions
    to the systematic uncertainty of the branching fraction
    measurements as explained in the text.}
  \label{tab:BSSum}
\end{table}
\clearpage 

\subsection{Improved Averages for $B(\tauKpn)$ and $B(\tauKpp)$}
\label{sub:ImpB}

\begin{figure}[!hb]
\begin{center}
  \includegraphics[width=.46\textwidth]{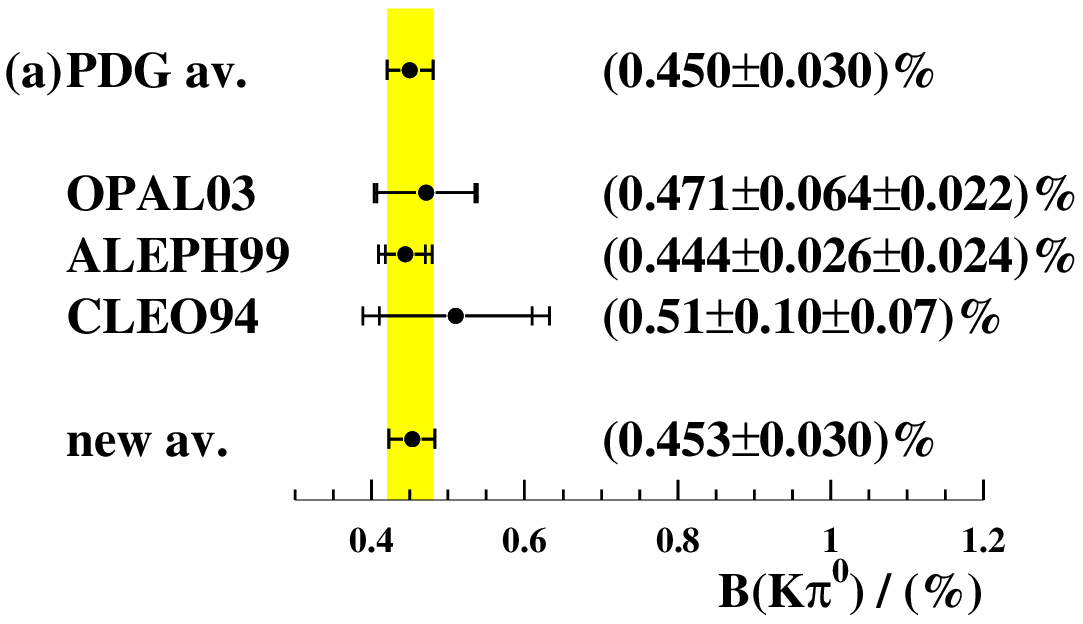}\hfill
  \includegraphics[width=.49\textwidth]{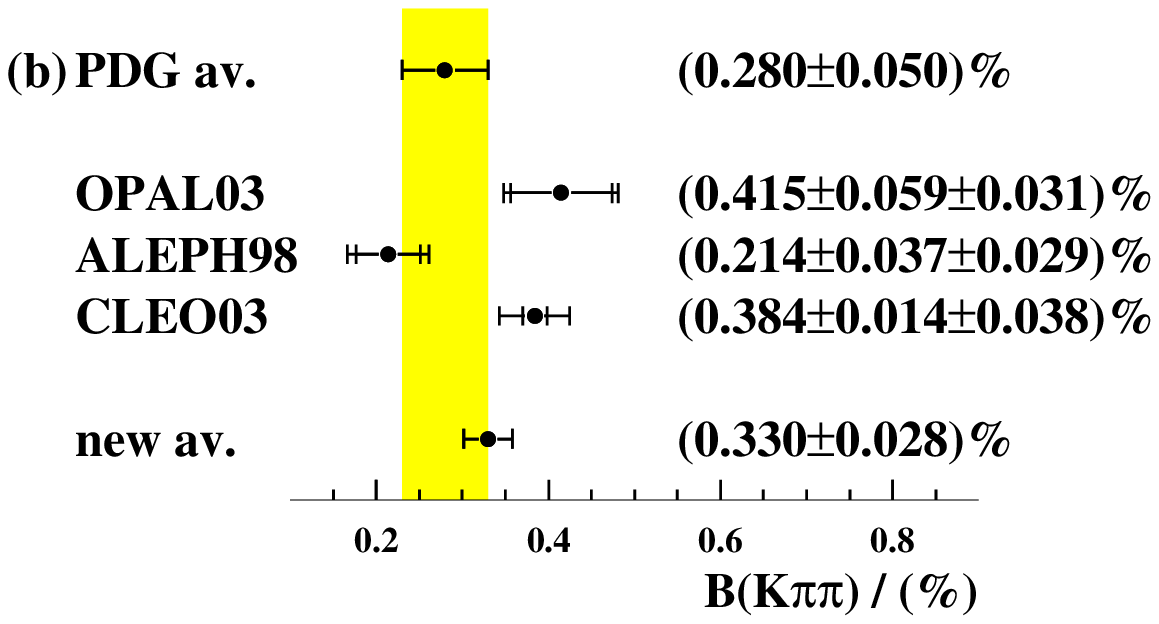}
  \caption{\em New averages for the branching fractions for $\tauKpn$ (a)
  and $\tauKpp$ (b). The previous world average as given by the PDG is
  indicated by the shaded band. The measurements used in the
  calculation of the weighted mean and the new averages are
  shown. OPAL03 denotes the values for the branching fractions
  obtained from this analysis.} 
  \label{fig:NewBAv}
\end{center}
\end{figure}

For the determination of the spectral function and its moments
described below, new average values for the branching fractions of the
decays $\tauKpn$ and $\tauKpp$ are determined. In addition to the
results obtained here, the same measurements are used as inputs for
the calculation as in \cite{bib:PDG}. For the channel $\tauKpp$, the
previous result from OPAL \cite{bib:SherryKpipi} was replaced by the
measurement obtained in this analysis. Also, the CLEO result was
updated using \cite{bib:CLEOKpp}.
The new averages are:
\begin{eqnarray*}
  B_\mathrm{av}(\tauKpn) &=& (0.453\pm0.030)\% \\
  B_\mathrm{av}(\tauKpp) &=& (0.330\pm0.028)\%. 
\end{eqnarray*}
The measurements used, together with the averages given in
\cite{bib:PDG} and the improved values for the branching fractions
obtained here are displayed in Figure \ref{fig:NewBAv}.   

The branching fraction obtained from this analysis for the \Kpn\
channel is consistent with the previous measurements within the errors
quoted. The value obtained for the \Kpp\ channel is consistent with
the new measurement from CLEO and the theoretical prediction in
\cite{bib:Finkemeier1996hh}. The ALEPH \cite{bib:ALEPHKpp} result
differs from these values by roughly $2.5\sigma$. 

\subsection{The Strangeness Spectral Function}\label{sub:SpFctn}
The hadronic decay of the \Pgt\ lepton is commonly written in terms of
the so-called spectral functions $v_1(s)$, $a_{0/1}(s)$ for the
non-strange part and $v_{0/1}^\mathrm{S}(s)$ and
$a_{0/1}^\mathrm{S}(s)$ for the strange part. The functions $v$ and
$a$ are the vector (V) and the axialvector (A) contributions,
respectively, while the subscript denotes the angular momentum
$J$. The variable $s$ is the invariant mass squared of the hadronic
system. The spectral function is experimentally determined by
measuring the invariant mass spectra of the given hadronic modes and
normalizing them to their respective branching fractions. The
contributions to the total strangeness spectral function then read:
\begin{equation}\label{eq:v+a1}
  v_1^\mathrm{S}(s)/a_1^\mathrm{S}(s) = \frac{m_\Pgt^2}{6 |V_\mathrm{us}|^2 S_\mathrm{ew}}
  \left(1-\frac{s}{m_\Pgt^2}\right)^{-2}\left( 1+\frac{2s}{m_\Pgt^2}\right)^{-1}
            \frac{B(\Pgt\to (V/ A)^{(S=-1,J=1)}\nu_\Pgt)}
                  {B(\Pgt\to e^-\bar\nu_e\nu_\Pgt)} 
             \frac{1}{N_\mathrm{V/A}}\frac{\mathrm{d}N_\mathrm{V/A}}{\mathrm{d}s}
\end{equation}
and
\begin{equation}\label{eq:v+a0}
  v_0^\mathrm{S}(s)/a_0^\mathrm{S}(s) =  \frac{m_\Pgt^2}{6 |V_\mathrm{us}|^2 S_\mathrm{ew}}
  \left(1-\frac{s}{m_\Pgt^2}\right)^{-2}
  \frac{B(\Pgt\to (V/ A)^{(S=-1,J=0)}\nu_\Pgt)}
                  {B(\Pgt\to e^-\bar\nu_e\nu_\Pgt)} 
             \frac{1}{N_\mathrm{V/A}}\frac{\mathrm{d}N_\mathrm{V/A}}{\mathrm{d}s}
\end{equation}
where $|V_\mathrm{us}|=0.2196\pm0.0023$ \cite{bib:PDG} is the CKM weak mixing
matrix element, $m_\Pgt=(1\,776.9^{+0.31}_{-0.27})\,\mathrm{MeV}$
\cite{bib:BES} and $S_\mathrm{ew}=1.0194\pm 0.0040$
\cite{bib:Marciano} is an electroweak correction factor. The total
strangeness spectral function $(v+a)$ is then obtained by adding the
individual contributions. To disentangle the vector and the
axialvector contributions for the spin-1 part, a detailed analysis of the
resonance structure of the measured spectra is necessary which is
not done here due to the limited statistics. The kaon pole contributes
to the pseudoscalar spin-0 part.

\begin{figure}[!b]
  \begin{center}
    \includegraphics[width=.7\textwidth]{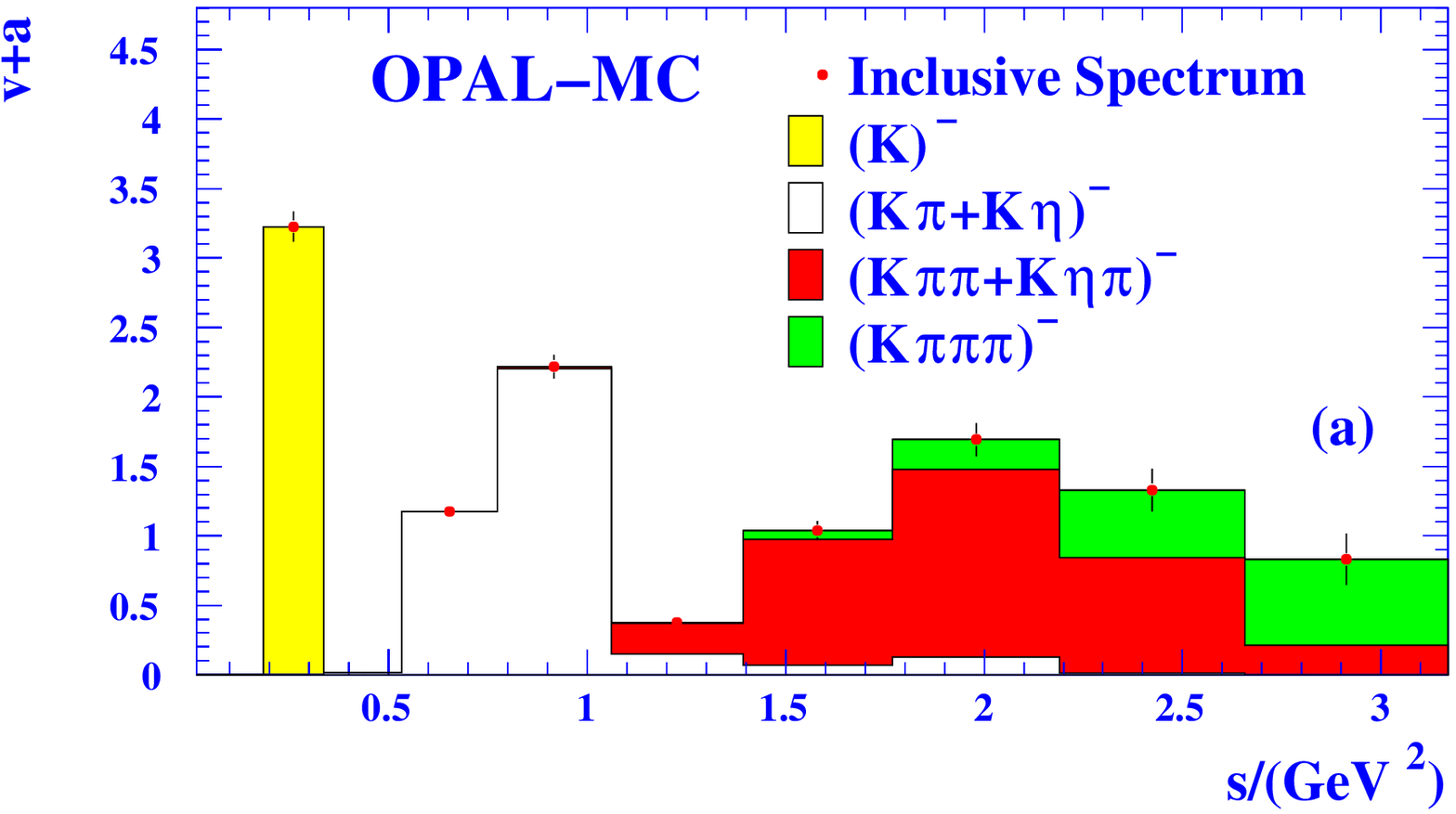}
    \includegraphics[width=.7\textwidth]{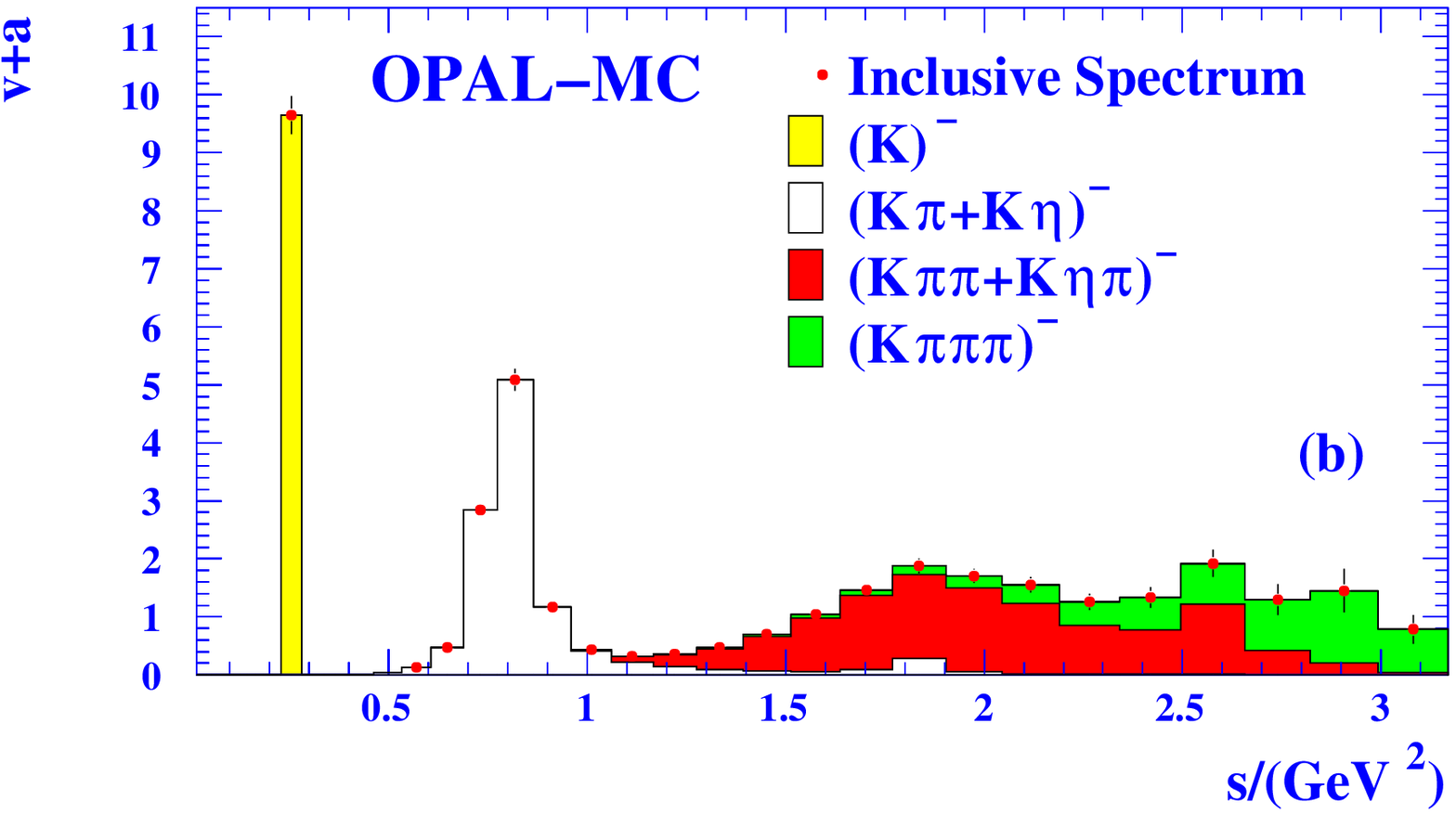}
    \caption{\em The strangeness spectral function as determined from Monte
    Carlo. For illustration purposes two
    different binnings were used -- (a) $150\,\mathrm{MeV}$ and (b)
    $50\,\mathrm{MeV}$ in mass. The dots represent the inclusive
    spectrum. The white histogram denotes the contribution from the $\Kpm$
    and $\Ketam$ final states, the dark shaded histogram those from the
    $\Kppm$ and $\Ketapm$ channels and the light shaded area shows the
    contribution from $\Kpppm$ final states. The leftmost bin
    represents the kaon pole which is the only contribution to the
    spin-0 part. The errors shown include the statistical uncertainty
    of the Monte Carlo as well as the uncertainties on the branching
    fractions.} 
    \label{fig:specfuncmc}
  \end{center}
\end{figure}

Summing the individual contributions (see Equations \ref{eq:v+a1} and
\ref{eq:v+a0}), the strangeness spectral function is obtained. The
Monte Carlo prediction of the total strangeness spectral function as a
function of the invariant mass squared is displayed in Figure
\ref{fig:specfuncmc}. The improved version of the \Pgt\ Monte Carlo as
explained in Section \ref{sub:Smlt} has been used here. The branching
fractions as given in Table \ref{tab:channels} have been used to weight
the individual mass spectra. For illustration purposes the spectral
function is shown using two different binnings. A non-equidistant
binning is chosen which corresponds to a bin width of
$50\,\mathrm{MeV}$ and $150\,\mathrm{MeV}$ in the invariant mass,
respectively. The errors given here are from Monte Carlo statistics
only.   

\begin{figure}[!h]
  \begin{center}
    \includegraphics[width=.95\textwidth]{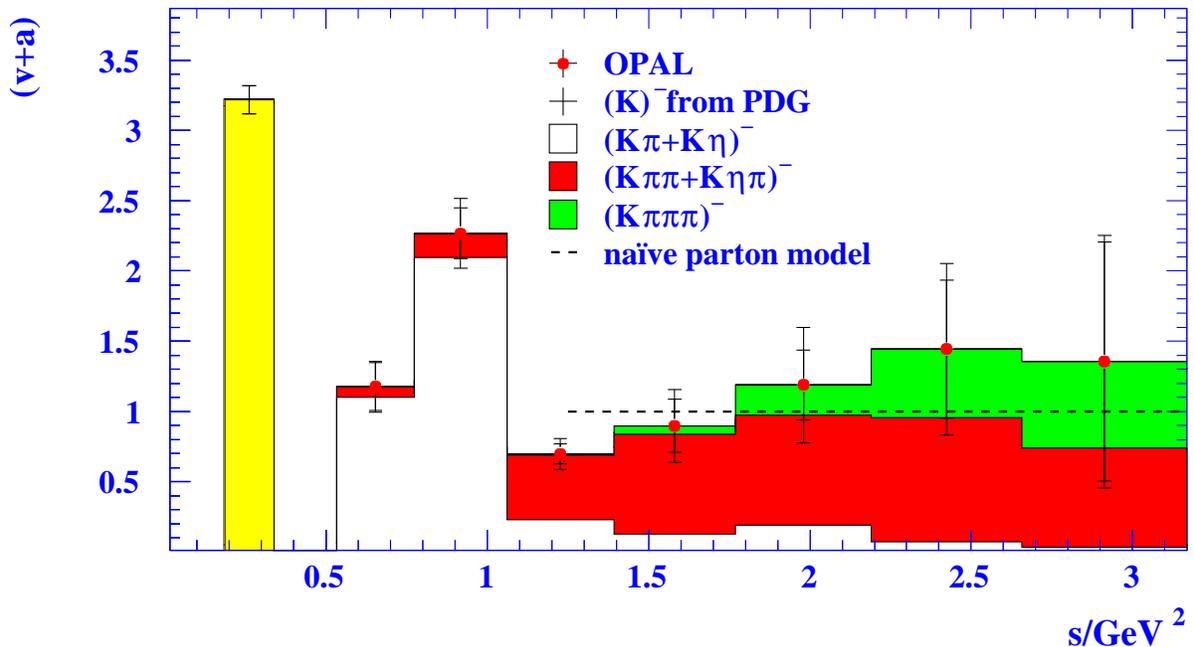}
    \caption{\em The spectral function from strange \Pgt\ lepton
      decays. The dots show the inclusive spectrum as measured in
      OPAL. The white histogram shows the exclusive $\KpKetam$
      spectrum, the dark shaded area the $\KppKetapm$ contribution and
      the light shaded area the contribution from the $\Kpppm$ final
      states. The error bars show the statistical uncertainty (inner
      error bar) and the systematic uncertainty added in quadrature
      (total error).} 
    \label{fig:specfunc}
  \end{center}
\end{figure}

The spectral function obtained from the data is displayed in Figure
\ref{fig:specfunc}. For $\tauKpn$ and $\tauKpp$, the new average branching
fractions and their respective errors as given in Section
\ref{sub:ImpB} are used. The binning chosen in this plot corresponds
to a bin width of $150\,\mathrm{MeV}$ in the invariant mass and is
governed by the mass resolution of the \KOppn\ final state. The dots
with error bars represent the inclusive spectrum. The inner error bars
are the statistical uncertainties. They include the uncertainty on the
efficiency and on Monte Carlo statistics. The total error is
calculated by adding up the statistical and systematic uncertainties
(explained in the next section) in quadrature. The numerical values
are given in Table \ref{tab:StrangeRes}. The systematic uncertainty is
dominated by the uncertainty on the \Pgt\ branching fractions.

For the $\Kpm$ final state both \Pgt\ decay channels
\KOp\ and \Kpn\ are measured. The channel \Keta\ which also
contributes to the two meson final state is taken from Monte
Carlo. For the $\Kppm$ final state the spectra \KOppn\ and \Kpp\ are
measured. The contribution from the decay \Kpnpn\ is added from Monte
Carlo as well as the \Ketap\ channel which also contributes to the
three meson final states. For the $\Kpppm$ spectrum, which consists of
the channels \Kpppn, \pKOpp, \Kpnpnpn\  and \KOppp, the prediction
from the Monte Carlo is taken. 

\subsection{Systematic Uncertainties on the Spectral Function}\label{sub:Syst}
The sources for possible systematic uncertainties listed below have been
considered. Since the individual contributions are different for the different
final states, the error is given for each bin in $s$ separately in
Table \ref{tab:StrangeRes}. 

\begin{itemize}
\item Energy loss measurement ($\Delta_{\dedx}$):\\
  The systematic variation is described in Section \ref{sub:BF}. In
  addition, the momentum dependence of this variation has been
  investigated. No significant influence has been found. 
\item Energy scale in \Pgpz\ reconstruction ($\Delta_\mathrm{E}$):\\
  The systematic variation is described in Section \ref{sub:BF}.
\item Momentum scale ($\Delta_\mathrm{p}$):\\
  The systematic variation is described in Section \ref{sub:BF}.
\item PDG errors on branching fractions ($\Delta_\mathrm{B}$):\\
  The dominant contribution to the systematic uncertainty comes from the
  uncertainty in the \Pgt\ branching fractions (see Table
  \ref{tab:channels}). The branching fractions are varied randomly and
  the difference in the result is quoted as systematic
  uncertainty. The channels which populate the region of
  high $s$ have branching fractions with relative errors close to
  $100\%$ leading to large uncertainty in the spectral function
  itself. This also covers the uncertainties on the shape.
\item \PKzS\ identification ($\Delta_\mathrm{\PKzS}$):\\
  A possible origin for systematic effects in the \PKzS\ identification is
  the estimation of the background using Monte Carlo. In particular the
  number of photon conversions found in \Pgt\ lepton decays is not perfectly
  modeled. The cut on the $\chi^2$ probability of the 2C constrained
  fit to \PKzS\ (see Section \ref{sub:IKOs}) has been varied from
  $10^{-5}$ to 0.01 to estimate a possible systematic effect. This cut reduces
  the number photon conversions in the sample by one order of
  magnitude. 
\item Mass correction procedure ($\Delta_\mathrm{mcorr}$):\\
  In order to assess the systematic uncertainty associated with the
  mass correction procedure, two possible sources have to be
  considered. The effect due to different input mass spectra and the
  effect due to the choice of the number of iterations. They have been
  studied using high statistics event samples. The systematic
  deviations using either flat, phase space, or resonance shaped input
  spectra (with an optimized number of iterations for each scenario)
  are about $5\%$ in each mass bin. The deviations are largest when a
  flat input distribution is assumed. Because the true mass spectrum
  is certainly not flat, we consider $5\%$ to be a conservative
  estimate for $\Delta_\mathrm{mcorr}$. The effect of the number
  of iterations was estimated using a phase space distribution for 
  input spectrum and performing one additional iteration step. The
  change was found to be negligible compared to the effect of varying 
  the input distribution.  
\end{itemize}

\begin{table}[!h]
  \begin{center}
  \begin{tabular}{|c||c|c|c|c|c|c||cc||c|}\hline
    \both ($s$-range)/$\mathrm{GeV}^2$ 
    & $\Delta_\mathrm{B}$ 
    & $\Delta_{\mbox{d}E/\mbox{d}x}$ 
    & $\Delta_{\PKzS}$ 
    & $\Delta_\mathrm{E}$  
    & $\Delta_\mathrm{p}$  
    & $\Delta_\mathrm{mcorr}$
    & $\Delta_\mathrm{sys}^\mathrm{tot}$
    & $\Delta_\mathrm{stat}$
    & $v+a$
    \\\hline\hline
    \up   (0.53,\,0.77) & 0.04 & 0.006 & 0.006 & 0.007 & 0.003 & 0.06 & 0.07 & 0.17 & 1.17$\pm$ 0.18 \\
          (0.77,\,1.06) & 0.13 & 0.011 & 0.011 & 0.014 & 0.001 & 0.11 & 0.17 & 0.18 & 2.27$\pm$ 0.25 \\
          (1.06,\,1.39) & 0.08 & 0.003 & 0.003 & 0.004 & 0.001 & 0.03 & 0.09 & 0.07 & 0.69$\pm$ 0.11 \\
          (1.39,\,1.77) & 0.18 & 0.005 & 0.005 & 0.005 & 0.002 & 0.05 & 0.18 & 0.19 & 0.90$\pm$ 0.26 \\
          (1.77,\,2.19) & 0.32 & 0.006 & 0.007 & 0.007 & 0.003 & 0.06 & 0.33 & 0.25 & 1.22$\pm$ 0.41 \\
          (2.19,\,2.66) & 0.35 & 0.007 & 0.009 & 0.009 & 0.003 & 0.07 & 0.36 & 0.49 & 1.44$\pm$ 0.61 \\
    \down (2.66,\,3.17) & 0.30 & 0.007 & 0.008 & 0.008 & 0.003 & 0.07 & 0.31 & 0.85 & 1.35$\pm$ 0.90 \\\hline
  \end{tabular}
    \caption[Result of the analysis]
            {\em Result for the strangeness spectral function. The
            table shows the values of the total strangeness spectral
            function $(v+a)$ and the statistical and the total systematic
            uncertainties for every bin in $s$. The individual
            contributions to the systematic uncertainty are discussed
            in the text. The total uncertainty quoted is the quadratic
            sum of the statistical and systematic uncertainties.}    
    \label{tab:StrangeRes}
  \end{center}
\end{table}

\subsection{The Spectral Moments $R^{kl}_{\Pgt,S}$}\label{sub:Mom}
\enlargethispage*{.5cm}
The moments of the spectral function are defined as:
\begin{equation}
  R^{kl}_{\Pgt,S}(m_\Pgt^2) =\int_0^{m_\Pgt^2} \mathrm{d}s 
  \left( 1-\frac{s}{m_\Pgt^2}\right)^k
  \left( \frac{s}{m_\Pgt^2}\right)^l
  \frac{B(\Pgt\to (V/ A)^{(S=-1,J=0/1)}\nu_\Pgt)}
       {B(\Pgt\to e^-\bar\nu_e\nu_\Pgt)} 
  \frac{1}{N_\mathrm{V/A}}\frac{\mathrm{d}N_\mathrm{V/A}}{\mathrm{d}s},
\end{equation}
where the sum runs over all final states with open strangeness.
The values measured for the moments
$kl=(00,\,10,\,11,\,12,\,13,\,20,\,21,\,30,\,40)$, their statistical
and systematic uncertainties are given in Table \ref{tab:MomS}. The
various sources contributing to the systematic uncertainty have been
discussed in Section \ref{sub:Syst}. The value for $R_{\Pgt,S}^{00}$
is calculated from the branching fractions only and is therefore
independent of the measured spectra.  
In addition, the CKM-weighted difference of the corresponding strange and
non-strange moments is given:
\begin{equation}
\delta R^{kl}_{\Pgt} 
  =  \frac{R^{kl}_{\Pgt,non-S}}{|V_\mathrm{ud}|^2} 
   - \frac{R^{kl}_{\Pgt,S    }}{|V_\mathrm{us}|^2},
\end{equation}
(see Table \ref{tab:MomS}) where $R^{kl}_{\Pgt,S}$ are the strange
moments and $R^{kl}_{\Pgt,non-S}$ is the
sum of the vector- and axialvector non-strange moments from 
\cite{bib:OPALSven}, using updated branching fractions. The values of
the matrix elements used for the non-strange and strange moments are
$|V_\mathrm{ud}|=(0.9734\pm0.0008)$ and
$|V_\mathrm{us}|=(0.2196\pm0.0023)$, respectively
\cite{bib:PDG}. Systematic errors which are common to both, strange and
non-strange moments, such as the energy scale error and the
momentum scale for tracks, are $100\%$ correlated and are treated
accordingly.  
\clearpage

\begin{sidewaystable}
  \centering\small
  \begin{tabular}{|r|c@{$\,\pm\,$}c||ccccccccc|}\hline
    \both $kl$ & \multicolumn{2}{c||}{$R^{kl}_{\Pgt,\mathrm{non-S}}$} 
                               & $00$ & $10$ & $11$ & $12$ & $13$ & $20$ & $21$ & $30$ & $40$ \\\hline 
    \up   $00$ & 3.469 & 0.014 & 100 &     &     &     &     &     &     &     &     \\
          $10$ & 2.493 & 0.013 &  66 & 100 &     &     &     &     &     &     &     \\
          $11$ & 0.549 & 0.004 &  68 &  65 & 100 &     &     &     &     &     &     \\
          $12$ & 0.203 & 0.002 &  51 &   9 &  74 & 100 &     &     &     &     &     \\
          $13$ & 0.092 & 0.002 &  33 & -26 &  33 &  86 & 100 &     &     &     &     \\
          $20$ & 1.944 & 0.011 &  55 &  93 &  45 & -11 & -40 & 100 &     &     &     \\
          $21$ & 0.346 & 0.003 &  59 &  86 &  88 &  35 & -13 &  71 & 100 &     &     \\
          $30$ & 1.597 & 0.009 &  48 &  85 &  28 & -24 & -44 &  93 &  58 & 100 &     \\
    \down $40$ & 1.362 & 0.008 &  42 &  77 &  14 & -30 & -43 &  87 &  43 &  92 & 100 \\\hline
  \end{tabular}\\[1ex]
  \begin{tabular}{|r|c@{$\,\pm\,$}c||c|ccccc||ccccccccc|}\hline
    \both $kl$ & \multicolumn{2}{c||}{$R^{kl}_{\Pgt,\mathrm{S}}$} 
                               & $\Delta_\mathrm{stat}$                                
                               & $\Delta_{\dedx}$ 
                               & $\Delta_{\PKzS}$ 
                               & $\Delta_\mathrm{E}$  
                               & $\Delta_\mathrm{p}$  
                               & $\Delta_\mathrm{mcorr}$
                               & $00$ & $10$ & $11$ & $12$ & $13$ & $20$ & $21$ & $30$ & $40$ \\\hline 
    \up   $00$ & 0.1677 & 0.0050 & -- & -- & -- & -- & -- & --                         & 100 &      &     &     &     &     &     &     &     \\
          $10$ & 0.1161 & 0.0038 & 0.0035 & 0.0006 & 0.0006 & 0.0005 & 0.0002 & 0.0011 &  89 &  100 &     &     &     &     &     &     &     \\
          $11$ & 0.0298 & 0.0012 & 0.0011 & 0.0001 & 0.0001 & 0.0001 & 0.0001 & 0.0004 &  97 &   83 & 100 &     &     &     &     &     &     \\
          $12$ & 0.0107 & 0.0006 & 0.0005 & 0.0002 & 0.0002 & 0.0002 & 0.0001 & 0.0002 &  86 &   54 &  91 & 100 &     &     &     &     &     \\
          $13$ & 0.0048 & 0.0004 & 0.0002 & 0.0002 & 0.0002 & 0.0002 & 0.0001 & 0.0001 &  74 &   36 &  78 &  97 & 100 &     &     &     &     \\
          $20$ & 0.0862 & 0.0028 & 0.0025 & 0.0006 & 0.0006 & 0.0006 & 0.0002 & 0.0008 &  75 &   97 &  66 &  32 &  13 & 100 &     &     &     \\
          $21$ & 0.0191 & 0.0007 & 0.0006 & 0.0001 & 0.0001 & 0.0001 & 0.0001 & 0.0002 &  92 &   96 &  92 &  66 &  47 &  87 & 100 &     &     \\
          $30$ & 0.0671 & 0.0022 & 0.0020 & 0.0005 & 0.0005 & 0.0004 & 0.0002 & 0.0006 &  66 &   92 &  54 &  19 &   1 &  99 &  78 & 100 &     \\
    \down $40$ & 0.0539 & 0.0018 & 0.0016 & 0.0003 & 0.0003 & 0.0003 & 0.0001 & 0.0005 &  60 &   87 &  46 &  11 &  -4 &  96 &  70 &  99 & 100 \\\hline
  \end{tabular}\\[1ex]
  \begin{tabular}{|r|c@{$\,\pm\,$}c||ccccccccc|}\hline
    \both $kl$ & \multicolumn{2}{c||}{$\delta R^{kl}_\Pgt$} 
                               & $00$ & $10$ & $11$ & $12$ & $13$ & $20$ & $21$ & $30$ & $40$ \\\hline 
    \up   $00$ &  0.184 & 0.128 & 100 &     &     &     &     &     &     &     &     \\
          $10$ &  0.224 & 0.095 &  79 & 100 &     &     &     &     &     &     &     \\
          $11$ & -0.039 & 0.028 &  88 &  67 & 100 &     &     &     &     &     &     \\
          $12$ & -0.008 & 0.014 &  67 &  37 &  65 & 100 &     &     &     &     &     \\
          $13$ & -0.002 & 0.009 &  49 &  20 &  48 &  53 & 100 &     &     &     &     \\
          $20$ &  0.264 & 0.070 &  64 &  74 &  51 &  20 &   5 & 100 &     &     &     \\
          $21$ & -0.031 & 0.017 &  81 &  76 &  75 &  46 &  26 &  66 & 100 &     &     \\
          $30$ &  0.294 & 0.055 &  56 &  71 &  42 &  11 &   0 &  73 &  60 & 100 &     \\
    \down $40$ &  0.320 & 0.045 &  52 &  69 &  36 &   6 &  -4 &  74 &  55 &  77 & 100 \\\hline
  \end{tabular}\hfill
  \begin{tabular}{|r|c@{$\,\pm\,$}c||ccccccccc|}\hline
    \both $kl$ & \multicolumn{2}{c||}{$R_{\Pgt,S}^{kl}/R_{\Pgt,non-S}^{kl}$} 
                               & $00$ & $10$ & $11$ & $12$ & $13$ & $20$ & $21$ & $30$ & $40$ \\\hline 
    \up   $00$ & 0.0484 & 0.0015 & 100 &     &     &     &     &     &     &     &     \\
          $10$ & 0.0466 & 0.0015 &  79 & 100 &     &     &     &     &     &     &     \\
          $11$ & 0.0543 & 0.0022 &  88 &  67 & 100 &     &     &     &     &     &     \\
          $12$ & 0.0527 & 0.0030 &  67 &  37 &  66 & 100 &     &     &     &     &     \\
          $13$ & 0.0518 & 0.0045 &  49 &  20 &  48 &  53 & 100 &     &     &     &     \\
          $20$ & 0.0444 & 0.0015 &  64 &  74 &  51 &  20 &   5 & 100 &     &     &     \\
          $21$ & 0.0552 & 0.0021 &  80 &  76 &  75 &  46 &  26 &  66 & 100 &     &     \\
          $30$ & 0.0420 & 0.0014 &  57 &  71 &  42 &  12 &   0 &  73 &  59 & 100 &     \\
    \down $40$ & 0.0400 & 0.0013 &  53 &  69 &  36 &   7 &  -4 &  73 &  55 &  76 & 100 \\\hline
  \end{tabular}
  \caption{\it The spectral moments for $kl=(00, 10, 11, 12,
    13, 20, 21, 30, 40)$. The tables include the values for the
    non-strange moments, the strange moments, the weighted difference
    and the quotient of strange to non-strange moments. The values for
    the non-strange moments given are calculated from
    \cite{bib:OPALSven} using updated branching fractions. For the
    strange moments, statistical and systematic uncertainties are
    listed separately for the individual sources. The statistical
    uncertainty also contains the uncertainty on the branching
    fractions. The uncertainty on $R_\Pgt^{00}$ is the total error
    from the branching fractions as given in \cite{bib:PDG}. On the
    right hand side of each table, the correlations are given in percent.}  
  \label{tab:MomS}
\end{sidewaystable}
\clearpage

\section{Conclusions}\label{sec:Summary}
\enlargethispage*{.5cm}
The spectral function in \Pgt\ lepton decays into vector and
axial vector final states with open strangeness, as well as the spectral
moments $R_{\Pgt,S}^{kl}$, have been measured with the OPAL detector,
operating on the \PZz\ resonance. The selection of the decay channels 
requires a detailed study of the charged kaon identification using
\dedx, neutral pion reconstruction in the electromagnetic calorimeter, and
identification of \PKzS. For the decays \Pgtm $\rightarrow$ $\Kpm \Pgngt$,
$\Kppm  \Pgngt$ and $\Kpppm  \Pgngt$, $94\%$ of all final states have
been experimentally accessed; for the remaining $6.6\%$, as well as for 
the strange final states including \Pgh\ mesons, Monte Carlo simulations have
been used. The reconstructed strange final states corrected for
resolution effects and detection efficiencies yield the strangeness
spectral function of the \Pgt\ lepton.
The spectral moments $R_{\Pgt,S}^{kl}$ for
$kl=\{10,\,11,\,12,\,13,\,20,\,21,\,30,\,40\}$, as well as the ratio
of strange to non-strange moments, 
updated from \cite{bib:OPALSven}, and the weighted differences $\delta
R_{\Pgt}^{kl}$ are given, which are useful quantities for further
theoretical analysis.  For the decay channels $\tauKpn$ and
$\tauKpp$ competitive branching fractions are obtained:
\begin{eqnarray*}
  B(\tauKpn) &=& (0.471\pm0.064_\mathrm{stat}\pm0.022_\mathrm{sys})\% \\
  B(\tauKpp) &=& (0.415\pm0.059_\mathrm{stat}\pm0.031_\mathrm{sys})\%.
\end{eqnarray*}
\newpage

\par
Acknowledgements:
\par
We particularly wish to thank the SL Division for the efficient operation
of the LEP accelerator at all energies
 and for their close cooperation with
our experimental group.  In addition to the support staff at our own
institutions we are pleased to acknowledge the  \\
Department of Energy, USA, \\
National Science Foundation, USA, \\
Particle Physics and Astronomy Research Council, UK, \\
Natural Sciences and Engineering Research Council, Canada, \\
Israel Science Foundation, administered by the Israel
Academy of Science and Humanities, \\
Benoziyo Center for High Energy Physics,\\
Japanese Ministry of Education, Culture, Sports, Science and
Technology (MEXT) and a grant under the MEXT International
Science Research Program,\\
Japanese Society for the Promotion of Science (JSPS),\\
German Israeli Bi-national Science Foundation (GIF), \\
Bundesministerium f\"ur Bildung und Forschung, Germany, \\
National Research Council of Canada, \\
Hungarian Foundation for Scientific Research, OTKA T-038240, 
and T-042864,\\
The NWO/NATO Fund for Scientific Research, the Netherlands.\\



\begin{thebibliography}{99}
%
%
\bibitem{bib:Braaten}
  E.~Braaten, S.~Narison and A.~Pich,
  Nucl.~Phys. {\bf B373}, (1992) 581.
%
%
\bibitem{bib:Braaten88}
  E.~Braaten,
  Phys.~Rev.~Lett. {\bf 60}, (1988) 1606.
%
%
\bibitem{bib:Braaten89}
  E.~Braaten,
  Phys.~Rev. {\bf D39}, (1989) 1458.
%
%
\bibitem{bib:Narison88}
  S.~Narison and A.~Pich,
  Phys.~Lett. {\bf B211}, (1988) 183.
%
%
\bibitem{bib:Pich92}
  F.~{Le~Diberder} and A.~Pich,
  Phys.~Lett. {\bf B286}, (1992) 147.
%
%
\bibitem{bib:Pivovarov1991rh}
A.~A.~Pivovarov,
Z.\ Phys.\ C {\bf 53}, (1992) 461.
%
%
\bibitem{bib:OPALSven}
  OPAL Collaboration, K.~Ackerstaff {\it et al.},  
  Eur.~Phys.~J. {\bf C7}, (1999) 571.
%
%
\bibitem{bib:ALEPHQCD}
  ALEPH Collaboration, R.~Barate {\it et al.},
  Eur.~Phys.~J. {\bf C4}, (1998) 409.
%
%
\bibitem{bib:CLEOQCD}
  CLEO Collaboration, T.~Coan {\it et al.},
  Phys.~Lett. {\bf B356}, (1995) 580.
%
%
\bibitem{bib:TheoAlphas}
  See for example A.~Pich,
  `QCD predictions for the tau hadronic width: Determination of 
   {$\as(m_\tau)$}' in {\it Proceedings of the 2nd Workshop on Tau Lepton
   Physics}, edited by K.~K.~Ghan, Columbus, Ohio, 1992.
%
%
\bibitem{bib:Chetyrkin}
K.~G.~Chetyrkin, J.~H.~Kuhn and A.~A.~Pivovarov,
Nucl.\ Phys.\ B {\bf 533}, (1998) 473.
%
%
\bibitem{bib:Pich1999hc}
A.~Pich and J.~Prades,
JHEP {\bf 9910}, (1999) 004.
%
%
\bibitem{bib:Maltman2002wb}
K.~Maltman,
eConf {\bf C0209101}, (2002) WE05.
%
%
\bibitem{bib:PDG} 
  K. Hagiwara et al., Phys. Rev. {\bf D 66}, (2002) 010001 (URL:
  http://pdg.lbl.gov/2003/s035.pdf). 
%
%
\bibitem{bib:smjt}
  OPAL Collaboration, G.~Abbiendi {\it et al.},
  Eur.~Phys.~J. {\bf C13}, (2000) 197.
%
%
\bibitem{bib:TPsel}
  OPAL Collaboration, R.~Akers et~al.,
  Phys.~Lett. {\bf B328}, (1994) 207.
%
%
\bibitem{bib:Detector} 
  OPAL Collaboration, K.~Ahmet {\it et al.},
  Nucl.~Instr.~and Meth. {\bf A 305}, (1991) 275.
%
%
\bibitem{bib:DetectorSI} 
  P.P.~Allport {\it et al.}, 
  Nucl.~Instr.~and Meth. {\bf A 324}, (1993) 34; \\
  P.P.~Allport {\it et al.}, 
  Nucl.~Instr.~and Meth. {\bf A 346}, (1994) 476.
%
%
\bibitem{bib:KoralZ}
  S.~Jadach, {B.F.L. Ward}, and Z.~W\c{a}s,
  Comp.~Phys.~Comm. {\bf 79}, (1994) 503.
%
%
\bibitem{bib:Tauola}
  S.~Jadach, Z.~W\c{a}s, R.~Decker, and {J.H. K\"{u}hn},
  Comp.~Phys.~Comm. {\bf 76}, (1993) 361.
%
%
\bibitem{bib:Finkemeier1995sr}
  M.~Finkemeier and E.~Mirkes,
  Z.\ Phys.\ C {\bf 69}, (1996) 243.
%
%
\bibitem{bib:Finkemeier1996hh}
  M.~Finkemeier, J.~H.~K\"uhn and E.~Mirkes,
  Nucl.\ Phys.\ Proc.\ Suppl.\  {\bf 55C}, (1997) 169.
%
%
\bibitem{bib:GOPAL}
  J.~Allison et~al.,
  Nucl.~Instrum.~Methods {\bf A317}, (1992) 47.
%
%
\bibitem{bib:Jetset}
  T.~Sj\"{o}strand,
  Comp.~Phys.~Comm. {\bf 82}, (1994) 74.
%
%
\bibitem{bib:Bhwide}
  S.~Jadach, W.~Placzek and B.~F.~Ward,
  Phys.\ Lett.\ B {\bf 390}, (1997) 298.
%
%
\bibitem{bib:PHOJET}
  R.~Engel, and J.~Ranft
  Phys.~Rev. {\bf D54}, (1996) 4244.
%
%
\bibitem{bib:TWOGEN}
  A.~Buijs, W.~G.~Langeveld, M.~H.~Lehto and D.~J.~Miller,
  Comput.\ Phys.\ Commun.\  {\bf 79}, (1994) 523.
%
%
\bibitem{bib:Vermaseren}
  R.~Battacharya, J.~Smith, and G.~Grammer,
  Phys.~Rev. {\bf D15}, (1977) 3267.
%
%
\bibitem{bib:Vermaseren2}
  J.~Smith, {J.A.M. Vermaseren}, and G.~Grammer,
  Phys.~Rev. {\bf D15}, (1977) 3280.
%
%
\bibitem{bib:OPALdEdx}
  M.~Hauschild {\it et al.},
  Nucl.~Inst.~Meth. {\bf A314}, (1992) 74.
%
%
\bibitem{bib:Bethke}
  S.~Bethke, Z.~Kunszt, D.E.~Soper and W.J.~Stirling:
  Nucl.~Phys.~{\bf B370}, (1992) 310,
  erratum ibid.~{\bf B523}, (1998) 681. 
%
%
\bibitem{bib:SherryKpipi}
OPAL Collaboration ,G.~Abbiendi {\it et al.},
Eur.\ Phys.\ J.\ C {\bf 13}, (2000) 197.
%
%
\bibitem{bib:CLEOKpp}
CLEO Collaboration, R.~A.~Briere {\it et al.},
Phys.~Rev.~Lett. {\bf 90}, (2003) 181802.
%
%
\bibitem{bib:ALEPHKpp}
ALEPH Collaboration, R.~Barate {\it et al.},
Eur.\ Phys.\ J.\ C {\bf 1}, (1998) 65.
%
%
\bibitem{bib:BES}
  BES Collaboration, J.Z.~Bai {\it et al.},
  Phys.~Rev.~{\bf D53}, (1996) 20.
%
%
\bibitem{bib:Marciano}
  {W.J. Marciano} and {A. Sirlin},
  Phys.~Rev.~Lett. {\bf 61}, (1988) 1815.
\end{thebibliography}
\end{document}